\documentstyle[12pt,psfig]{article}
\newcommand{\nsect}{\setcounter{equation}{0}
\def\theequation{\thesection.\arabic{equation}}\section}

\catcode`\@=11
\def\marginnote#1{}
\def\ifmath#1{\relax\ifmmode #1\else $#1$\fi}

\def\mstop{m_{\,\widetilde{t}}}

\def\bold#1{\setbox0=\hbox{$#1$}%
     \kern-.025em\copy0\kern-\wd0
     \kern.05em\copy0\kern-\wd0
     \kern-.025em\raise.0433em\box0 }

\def\GENITEM#1;#2{\par\vskip6pt \hangafter=0 \hangindent=#1
   \Textindent{$ #2$ }\ignorespaces}

\newcount\hour
\newcount\minute
\newtoks\amorpm
\hour=\time\divide\hour by60
\minute=\time{\multiply\hour by60 \global\advance\minute by-
\hour}
\edef\standardtime{{\ifnum\hour<12 \global\amorpm={am}%
    \else\global\amorpm={pm}\advance\hour by-12 \fi
    \ifnum\hour=0 \hour=12 \fi
    \number\hour:\ifnum\minute<100\fi\number\minute\the\amorpm}}
\edef\militarytime{\number\hour:\ifnum\minute<100\fi\number\minute}
\def\draftlabel#1{{\@bsphack\if@filesw {\let\thepage\relax
  \xdef\@gtempa{\write\@auxout{\string
    \newlabel{#1}{{\@currentlabel}{\thepage}}}}}\@gtempa
    \if@nobreak \ifvmode\nobreak\fi\fi\fi\@esphack}
     \gdef\@eqnlabel{#1}}
\def\@eqnlabel{}
\def\@vacuum{}
\def\draftmarginnote#1{\marginpar{\raggedright\scriptsize\tt#1}}
\def\draft{\oddsidemargin -.5truein
        \def\@oddfoot{\sl preliminary draft \hfil
        \rm\thepage\hfil\sl\today\quad\militarytime}
        \let\@evenfoot\@oddfoot \overfullrule 3pt
        \let\label=\draftlabel
        \let\marginnote=\draftmarginnote

\def\@eqnnum{(\theequation)\rlap{\kern\marginparsep\tt\@eqnlabel}%
\global\let\@eqnlabel\@vacuum}  }
\def\preprint{\twocolumn\sloppy\flushbottom\parindent 1em
        \leftmargini 2em\leftmarginv .5em\leftmarginvi .5em
        \oddsidemargin -.5in    \evensidemargin -.5in
        \columnsep 15mm \footheight 0pt
        \textwidth 250mmin      \topmargin  -.4in
        \headheight 12pt \topskip .4in
        \textheight 175mm
        \footskip 0pt

\def\@oddhead{\thepage\hfil\addtocounter{page}{1}\thepage}
        \let\@evenhead\@oddhead \def\@oddfoot{} \def\@evenfoot{}
}
\def\titlepage{\@restonecolfalse\if@twocolumn\@restonecoltrue\onecolumn
     \else \newpage \fi \thispagestyle{empty}\c@page\z@
        \def\thefootnote{\fnsymbol{footnote}} }
\def\endtitlepage{\if@restonecol\twocolumn \else  \fi
        \def\thefootnote{\arabic{footnote}}
        \setcounter{footnote}{0}}  
\catcode`@=12
\relax
\def\esphal{E_{\rm sph}}
\def\be{\begin{equation}}
\def\ee{\end{equation}}
\def\bea{\begin{eqnarray}}
\def\eea{\end{eqnarray}}
\def\simlt{\stackrel{<}{{}_\sim}}
\def\simgt{\stackrel{>}{{}_\sim}}

\def\NPB#1#2#3{{\it Nucl.~Phys.} {\bf{B#1}} (19#2) #3}
\def\PLB#1#2#3{{\it Phys.~Lett.} {\bf{B#1}} (19#2) #3}
\def\PRD#1#2#3{{\it Phys.~Rev.} {\bf{D#1}} (19#2) #3}
\def\PRL#1#2#3{{\it Phys.~Rev.~Lett.} {\bf{#1}} (19#2) #3}
\def\ZPC#1#2#3{{\it Z.~Phys.} {\bf C#1} (19#2) #3}
\def\PTP#1#2#3{{\it Prog.~Theor.~Phys.} {\bf#1}  (19#2) #3}
\def\MPLA#1#2#3{{\it Mod.~Phys.~Lett.} {\bf#1} (19#2) #3}
\def\PR#1#2#3{{\it Phys.~Rep.} {\bf#1} (19#2) #3}
\def\AP#1#2#3{{\it Ann.~Phys.} {\bf#1} (19#2) #3}
\def\RMP#1#2#3{{\it Rev.~Mod.~Phys.} {\bf#1} (19#2) #3}
\def\HPA#1#2#3{{\it Helv.~Phys.~Acta} {\bf#1} (19#2) #3}
\def\JETPL#1#2#3{{\it JETP~Lett.} {\bf#1} (19#2) #3}
\def\JPG#1#2#3{{\it J. Phys. G.}{\bf G#1}(19#2) #3}
\def\IJMPA#1#2#3{{\it Int.~J.~Mod.~Phys.} {\bf A#1} (19#2) #3}

\def\mst1{m_{\;\widetilde{t}_{1}}}

\def\st{\;\widetilde{t}}

\def\mst2{m_{\;\widetilde{t}_{2}}}
\def\mst12{m_{\;\widetilde{t}_{1,2}}}
\def\mstlr{m_{\;\widetilde{t}_{L,R}}}
\def\mstl{m_{\;\widetilde{t}_L}}
\def\mstr{m_{\;\widetilde{t}_R}}
\def\msb1{m_{\;\widetilde{b}_{1}}}
\def\msb2{m_{\;\widetilde{b}_{2}}}
\def\msb12{m_{\;\widetilde{b}_{1,2}}}

\def\mtilde2{\widetilde{m}^{2}}

\relax

\begin{document}
\topmargin-2.5cm

%
\begin{titlepage}
\begin{flushright}
IEM-FT-127/96 \\
hep--ph/9605387 \\
\end{flushright}
\vskip 0.3in
\begin{center}{\Large\bf SPHALERONS IN THE MSSM
\footnote{Work supported in part by the European Union
(contract CHRX/CT92-0004) and CICYT of Spain
(contract AEN95-0195).} }
\vskip .5in
{\bf J.M. Moreno, D.H. Oaknin} and {\bf M. Quir\'os} \\
\vskip.35in
Instituto de Estructura de la Materia, CSIC, Serrano
123, 28006-Madrid, Spain
\end{center}
\vskip1.3cm
\begin{center}
{\bf Abstract}
\end{center}
\begin{quote}
We construct the sphaleron solution, at zero and finite
temperature, in the Minimal Supersymmetric Standard Model as a
function of the supersymmetric parameters, including the leading
one-loop corrections to the effective potential in the presence of
the sphaleron. At zero temperature we have included
the one-loop radiative corrections, dominated by the
top/stop sector.  The sphaleron energy $E_{\rm MSSM}$ 
mainly depends on an effective Higgs mass
${\displaystyle m_h^{\rm eff}=\lim_{m_A\gg m_W} m_h}$, 
where $m_h$ is the lightest CP-even Higgs mass and $m_A$ the 
pseudoscalar mass. We have compared it with the Standard
Model result, with $m_h^{\rm SM}=m_h^{\rm eff}$, and found
small differences (1-2\%) in all cases.
At finite temperature we have included the one-loop
effective potential improved by daisy diagram resummation. The
sphaleron energy at the critical temperature can be encoded in
the temperature dependence of the vacuum expectation value of
the Higgs field with an error $\simlt 10$\%. The light stop
scenario has been re-examined and the existence of a window
where baryon asymmetry is not erased after the phase transition,
confirmed. Although large (low) values of $m_h$ ($m_A$) are
disfavoured by the strength of the phase transition, that window 
(along with LEP results) allows for
$m_h\simlt 80$ GeV, $m_A\simgt 110$ GeV and 
$A_t\simlt 0.4\; m_Q$.
\end{quote}
\vskip1.cm

\begin{flushleft}
IEM-FT-127/96\\
May 1996 \\
\end{flushleft}
\end{titlepage}
\setcounter{footnote}{0}
\setcounter{page}{0}
\newpage
%
\nsect{Introduction}
't Hooft observation~\cite{anomaly} that baryon, and lepton, number is not 
conserved at the quantum level in the Standard Model (SM) of electroweak
interactions, due to the presence of an SU(2) triangular
anomaly, opened up the exciting possibility of baryon number
non-conservation at low-energy. Even if the probability for 
$\Delta B\neq 0$ processes at zero temperature is exponentially
suppressed by a factor $\sim\exp(-16\pi^2/g^2)$, where $g$ is
the SU(2) gauge coupling, such suppression factor is absent at
high temperatures~\cite{Manton}-\cite{Dine1} 
and, more precisely, at temperatures of the
order of 100 GeV, triggering the hope
of generating the baryon asymmetry of the 
Universe~\cite{baryogenesis} at
the electroweak phase transition~\cite{reviews}.  

The existence of sphalerons, static and unstable solutions to
classical field equations in the SU(2) gauge theory with Higgs
fields in the fundamental representation, has been known since
long ago~\cite{sphalerons}. A key observation was that of
Klinkhamer and Manton~\cite{Manton} who interpreted the
sphaleron solution as a saddle point of the energy
barrier separating gauge-inequivalent classical vacua. Detailed
numerical calculations of the sphaleron solutions and energy in
the SM were subsequently examined by many 
authors~\cite{Manton,AKY,Yaffe}. The precise value of
the sphaleron energy is of the utmost importance for mechanisms
aiming to explain the generation of the baryon asymmetry of the
Universe~\cite{Kuzmin1}-\cite{Bochkarev}. In particular the
sphaleron transition rate for temperatures below the phase
transition temperature~\cite{Carson}
\be
\label{sphaltrans}
\Gamma_{\rm sph}\sim\exp\left\{\frac{\esphal(T)}{T}\right\} ,
\ee
along with the bound obtained by comparing
Eq.~(\ref{sphaltrans}) with the expansion rate of the Universe at
the temperature of the electroweak phase
transition~\cite{Shapo1,Shapo2}
\be
\label{sphalbound}
\frac{\esphal(T_b)}{T_b}\simgt 45
\ee
where $T_b$ denotes the transition, or bubble formation
temperature, allows to put upper bounds on the Higgs boson mass
and thus discard models of electroweak symmetry
breaking~\cite{Bochkarev} on the basis of baryon asymmetry
generation.  

Even though the SM has all the required properties for the
generation of the baryon asymmetry (CP violation, baryon number
violating processes and non-equilibrium processes generated by a
first-order phase transition) they are not strong enough to generate
the necessary amount of baryon asymmetry. In particular, the
CP-violating Cabibbo-Kobayashi-Maskawa phases strongly restrict
the possible baryon number
generation~\cite{Farrar}, though the latter
remains as a very controversial subject. Even disregarding this
point, the phase transition is not strongly first order enough and,
moreover, condition (\ref{sphalbound}) translates into an upper bound 
on the Higgs boson mass~\cite{Carrington,Bagnasco,Kajantie} 
which is well below the present experimental lower bound of 65 GeV.

Imposing the requirement of baryon asymmetry generation at the
electroweak phase transition translates into the requirement of
new physics at the weak scale. As a matter of fact, when
extending the SM at the weak scale and probing the baryogenesis
capability of the extended model the detailed calculation of the
sphaleron energy should be an essential piece of information.
It has been proved that there are extra sources
of CP violation, suitable for baryogenesis,
in minimal extensions of the SM, in particular
in the SM with a gauge singlet~\cite{Anderson,Dine1,Dine2} 
(with CP-violating non-renormalizable couplings
arising from integrated out extra fields),
and  in two-Higgs doublet models~\cite{Turok1}
(with extra phases in the Higgs sector), and that the phase
transition is strongly first order enough not to wipe out, after
the phase transition, the previously generated baryon 
asymmetry~\cite{Espinosa,Turok2} 
for values of the Higgs mass beyond the
experimental bounds. In both cases the sphaleron energy has been
computed, Refs.~\cite{Kastening1} and \cite{Kastening2}
respectively,  
and can support the baryogenesis achievements of the
corresponding models. However, the sphaleron solutions have been
evaluated in these models using the tree-level potential at zero
temperature, so that the sphaleron energy at the phase
transition temperature relies on the assumption of the scaling
law 
\be
\label{scaling}
\esphal(T)=\esphal(0)\frac{\langle\phi(T)\rangle}
{\langle\phi(0)\rangle}
\ee
which has only been proved to hold approximately in the SM
case~\cite{Braibant}, a similar calculation being absent in the
above extensions of the SM.

The possibility that new physics at the electroweak scale be
provided by the Minimal Supersymmetric Standard
Model (MSSM)~\cite{susy} is a very appealing one on the
theoretical front, since it can technically solve the 
hierarchy problem. 
The MSSM is also the natural candidate for an effective
theory from a more fundamental theory valid at the high scale
($\simlt M_{P\ell}$), 
as e.g. string theory. Finally LEP electroweak precision
measurements predict in the MSSM the
unification of gauge couplings at a scale $\sim 10^{16}$ GeV,
which is the only present `evidence' for Grand Unification and
new physics beyond the SM. All of that has generated a plethora
of experimental searches for new physics within the MSSM
framework in present and planned colliders~\cite{lep2,LHC}.
Exploring the capability of the MSSM of generating the observed
baryon asymmetry of the Universe at the electroweak phase
transition is therefore of the highest interest, mainly in order
to confront the experimental searches of supersymmetry with
possible regions of the space of supersymmetric parameters where
baryogenesis might be possibly produced. In this sense new
sources of CP violation are present in the 
MSSM~\cite{CPMSSM,SBCP}, which
can serve to overcome the strong SM suppression factors
mentioned above~\cite{Farrar}. On the other hand, the strength of the
first order phase transition has been extensively studied in the
MSSM~\cite{earlyMSSM}-\cite{Delepine}, 
and proved that the weak phase transition in the SM
(driven by the gauge coupling) can be strengthened in the MSSM
(driven by the top Yukawa coupling) in the presence of light
supersymmetric partners of the top quark (stops) and small
values of $\tan\beta$~\cite{CQW}. However a detailed study of the
sphaleron solutions and the sphaleron energy in the MSSM was
missing. 

In this paper we will construct the sphaleron solutions in the
MSSM at zero and finite temperature, including in all cases the
leading one-loop radiative corrections. The planning of this
paper is as follows: In Section 2 we will construct the
sphaleron solutions and the sphaleron energy in the MSSM at zero
temperature, including the full set of one-loop radiative
corrections to the effective potential in the presence of the
sphaleron Higgs. They are dominated by the top Yukawa 
coupling, and we will consistently neglect the bottom Yukawa 
coupling, a reasonable approximation for $\tan\beta\simlt 15$, which
are the values we will consider. In fact, as we will show, only
for $\tan\beta\simlt 3$ the baryogenesis scenario is feasible.
We have compared in every case the sphaleron energy in the MSSM,
with arbitrary sets of supersymmetric parameters 
predicting the supersymmetric and Higgs mass spectra,
with the sphaleron energy in the SM with an effective  Higgs mass. 
The deviations are typically $\simlt 1.5$\%. In Section 3 we
will construct the sphaleron solutions and energy in the MSSM at
finite temperature. We have included the one-loop effective
potential at finite temperature, in the presence of the sphaleron
Higgs, improved by the resummation of daisy diagrams.
We have checked that the approximation given by
Eq.~(\ref{scaling}) is accurate with an error less than a few
percent for the cases where the phase transition is weak.
For the cases studied in Section 4, where we have applied the results
of Sections 2 and 3 to the recently proposed case of light stops,
the error can be as large as $\sim 10$\%.
We have qualitatively confirmed the results of Ref.~\cite{CQW}.
In particular we have found a baryogenesis  window for
$\mstop\simlt m_t$, $m_h\simlt 80$ GeV, $\tan\beta\simlt 3$ and 
$m_A\simgt 110$ GeV, although not all the previous inequalities
can be simultaneously saturated. Finally Section 5 contains our
conclusions. 

We have adopted, throughout this paper, the approximation $g'=0$,
where $g'$ is the $U(1)_Y$ gauge coupling, in order to use a
spherically symmetric ansatz for the sphaleron. The corrections
of ${\cal O}(g'^{\; 2})$ to the sphaleron energy have been evaluated in
Ref.~\cite{Manton}, computed in Ref.~\cite{angle} for the SM
sphaleron, and proved to be negative and $\simlt 1$\%. Also we
have kept in the effective potential only one-loop corrections
in the presence of the sphaleron Higgs, which are
dominated by the top/stop sector and the top Yukawa coupling. We
have neglected one-loop radiative corrections in the presence of
the sphaleron-$W$. This procedure should provide a rather accurate
one-loop approximation because the leading finite temperature
${\cal O}(T^2)$ corrections to the effective potential in the
presence of the sphaleron-$W$ vanish~\footnote{This is
due to the fact that the sphaleron solution is non-zero only for
transverse degrees of freedom whose (magnetic) Debye mass
vanishes to all orders in perturbation theory~\cite{GPY}. This is
not the case for longitudinal degrees of freedom, so that a
non-vanishing component along $W_0$ would trigger, through the
one-loop electric mass, sizeable corrections.}. Subleading
finite temperature corrections to the effective potential are
suppressed by powers of $g^n$, $n>2$, and can therefore be
neglected as compared to those proportional to powers of the top
Yukawa coupling. The approximation of considering only radiative
corrections to the effective potential (or effective action) in
the presence of the sphaleron Higgs, and neglecting
those generated in the presence of the sphaleron-$W$,
is consistent only in cases where the former are dominated by a
large coupling (e.g. in our case by the top Yukawa coupling)
while the latter are always provided by  extra 
powers of the gauge coupling
constants. In other words, the approximation should  work in
theories where putting $g=0$ in radiative corrections is a good 
approximation. An example of this kind of theories is the MSSM, 
where radiative corrections
are dominated by the top Yukawa coupling. Unlike
the MSSM, the finite temperature radiative corrections in the SM
are dominated by the gauge coupling and making this
approximation would miss some terms similar to those
considered~\cite{Braibant}.  

\nsect{Sphalerons at zero temperature}

In this section we will compute the static unstable
solutions of classical equations of motion in the MSSM. 
As we have said in Section 1 we will work in the approximation
of taking $g'=0$ so that the $U(1)_Y$ gauge field $B_{\mu}$ can be
consistently set to zero. This will allow a spherically
symmetric ansatz, as follows. The lagrangian density for the
$SU(2)$ gauge fields $W_{\mu}$ and the Higgs system
\be
\label{higgses}
\begin{array}{rl}
H_1 = &{\displaystyle  \left[
\begin{array}{c}
H_1^0 \\
H_1^-
\end{array}
\right]  }\\
 & \\
H_2  = &  {\displaystyle \left[
\begin{array}{c}
H_2^+ \\
H_2^0
\end{array}
\right] }
\end{array}
\ee
is given by
\be
\label{lagrangian}
{\cal L}=-\frac{1}{4} W^a_{\mu\nu} W^{a\mu\nu}
+\left(D_{\mu}H_1\right)^{\dagger}\left(D^{\mu}H_1\right)
+\left(D_{\mu}H_2\right)^{\dagger}\left(D^{\mu}H_2\right)
-V_{\rm eff}(H_1,H_2)
\ee
where the $SU(2)$ field strength is defined as,
\be
\label{fstrength}
W^a_{\mu\nu}=\partial_{\mu}W^a_{\nu}-\partial_{\nu}W^a_{\mu}
+g\epsilon^{abc}W^b_{\mu}W^c_{\nu}
\ee
the covariant derivatives are
\be
\label{covder}
D_{\mu}\equiv\partial_{\mu}-i\; \frac{g}{2}\; W^a_{\mu}\sigma^a
\ee
$\sigma^a$ being the Pauli matrices, and $V_{\rm eff}$ is the
effective potential where we will consider one-loop corrections.

We can expand the effective potential (at zero temperature) as
\be
\label{descomp}
V_{\rm eff}=V_0(H_1,H_2)+V_1(H_1,H_2)+\cdots
\ee
where $V_0$ is the tree-level potential and $V_1$ contains
the one-loop radiative corrections. They are determined from the
supersymmetric structure of the MSSM, with superpotential
\be
\label{superp}
W=h_t Q_L\cdot H_2 U_L^c+\mu H_1\cdot H_2
\ee
and from the soft-breaking terms. In particular, the tree-level
potential is given by
\be
\label{tree}
\begin{array}{rl}
V_0(H_1,H_2) = & m_1^2\; H_1^{\dagger}H_1+
m_2^2\; H_2^{\dagger}H_2+m_3^2(H_1\cdot H_2+h.c.) \\ & \\
+&{\displaystyle  \frac{g^2}{8} } \left[
\left(H_1^{\dagger}H_1-H_2^{\dagger}H_2\right)^2 + 4
\left(H_1^{\dagger}H_2\right)\left(H_2^{\dagger}H_1\right)
\right]
\end{array}
\ee
and the one-loop corrections, in the 't Hooft-Landau gauge and
in the $\overline{\rm DR}$ renormalization scheme, by
\be
\label{oneloop}
V_1(H_1,H_2)=\sum_i\frac{n_i}{64\pi^2}m_i^4(H_1,H_2)
\left[\log\frac{m_i^2(H_1,H_2)}{Q^2}-\frac{3}{2}\right]
\ee
where $Q$ is the renormalization scale, that we are taking, for
definiteness, as $Q^2=m_t^2$, $m_i^2(H_1,H_2)$ is the field
dependent mass of the $i^{\rm th}$ particle in the background
$H_1$, $H_2$, and $n_i$ is the corresponding number of degrees
of freedom, which is taken negative for fermions. The relevant
degrees of freedom for our calculation are the gauge bosons
($W,\; Z$), the top quark ($t$) and its supersymmetric partners 
($\widetilde{t}_1$, $\widetilde{t}_2$), with
\be
\label{number}
n_t=-12,\ n_W=6,\ n_Z=3,\
n_{\;\widetilde{t}_1}=n_{\;\widetilde{t}_2}=6 .
\ee

Choosing the temporal, radial gauge
\be
\label{gauge}
W_0=0,\ \ x_i W_i^a=0\ (i=1-3,a=1-3)
\ee
we can write the static, spherically symmetric ansatz
\bea
\label{ansatz}
W^a_j(\vec{x})=\frac{2 f(r)}{g r^2}\epsilon_{ajk}x_k \nonumber \\
H_1(\vec{x})=h_1(r)\ i\; \frac{\vec{\sigma}\cdot \vec{x}}{r}
\left[
\begin{array}{c}
1 \\
0
\end{array}
\right] \\
H_2(\vec{x})=h_2(r)\ i\; \frac{\vec{\sigma}\cdot \vec{x}}{r}
\left[
\begin{array}{c}
0 \\
1
\end{array}
\right] \nonumber
\eea
for the sphaleron in the MSSM, where
$r^2\equiv x^2+y^2+z^2$. 

There, however, remain time-independent gauge transformations
which can be used to rotate the background fields in
(\ref{ansatz}) into a more convenient form for our calculation.
In particular, the gauge transformation~\cite{AKY}
\be
\label{gtransf}
U(\vec{x})=\exp\left[-i\ \frac{\pi}{2}\
\frac{\vec{\sigma}\cdot\vec{x}}{r}\right]
\ee
transforms the ansatz (\ref{ansatz}) into the background fields
\bea
\label{background}
W^a_j(\vec{x})=\frac{2 [1-f(r)]}{g r^2}\epsilon_{ajk}x_k \nonumber \\
H_1(\vec{x})=h_1(r)\
\left[
\begin{array}{c}
1 \\
0
\end{array}
\right] \\
H_2(\vec{x})=h_2(r)\ 
\left[
\begin{array}{c}
0 \\
1
\end{array}
\right] \nonumber
\eea
which should smoothly approach a vacuum in the unitary gauge as
$r\rightarrow \infty$.

In the presence of the background (\ref{background}) the
tree-level potential reads as
\be
\label{treefin}
V_0(h_1,h_2) =  m_1^2\; h_1^2(r)+
m_2^2\; h_2^2(r)+2\;m_3^2 h_1(r)h_2(r) 
+\frac{g^2}{8}  \left[h_1^2(r)-h_2^2(r)\right]^2
\ee
while the one-loop corrections are given by Eq.~(\ref{oneloop})
with  background dependent masses. 
The relevant masses are the top quark mass, given by
\be
\label{topmass}
m_t^2(h)=h_t^2 h_2^2(r),
\ee
the gauge boson masses, by
\be
\label{Wmass}
m_W^2(h)=m_Z^2(h)
=\frac{1}{2} g^2\left[h_1^2(r)+h_2^2(r)\right],
\ee
and the stop squared mass matrix given by 
\be
\label{stopmat}
{\cal M}^2_{\widetilde{t}}=\left(
\begin{array}{cc}
m_{\;\widetilde{t}_L}^2 & m_{\;\widetilde{t}_{LR}}^2 \\
m_{\;\widetilde{t}_{LR}}^2   & m_{\;\widetilde{t}_R}^2
\end{array}
\right)
\ee
with entries
\be
\begin{array}{rl}
\label{stopentr}
m_{\;\widetilde{t}_L}^2(h)= & m_Q^2+m_t^2(h)
+\frac{1}{4}g^2
\left[h_1^2(r)-h_2^2(r)\right]  \\ & \\
m_{\;\widetilde{t}_R}^2(h)= & m_U^2+m_t^2(h) \\ & \\
m_{\;\widetilde{t}_{LR}}^2(h)
= & h_t\left(A_t h_2(r)+\mu h_1(r)\right) ,
\end{array}
\ee
the mass eigenstates being defined by the diagonalization of
matrix (\ref{stopmat}),
\be
\label{stopeigen}
\mst12^2(h)=\frac{\mstl^2(h)+\mstr^2(h)}{2}\pm
\sqrt{\left[\frac{\mstl^2(h)-\mstr^2(h)}{2}\right]^2 
+\left[\mstlr^2(h)\right]^2 }
\ee
with $h\equiv(h_1,h_2)$.

By minimizing the effective potential (\ref{descomp}) with
respect to $h_1,h_2$, and imposing the minimum of the potential
at $(v_1,v_2)$, with $v=\sqrt{v_1^2+v_2^2}=171.1$ GeV, and 
$\tan\beta=v_2/v_1$ fixed, we can eliminate $m_1^2$ and $m_2^2$
in favour of the other parameters of the theory,
as~\cite{mariano2,ERZ} 
\be
\begin{array}{rl}
\label{m12}
m_1^2= & {\displaystyle
-m_3^2\tan\beta-\frac{m_Z^2}{2}\cos 2\beta
-\sum_i\frac{n_i}{64\pi^2}
\left[ \frac{\partial m_i^2}{\partial h_1}\frac{m_i^2}{h_1}
\left(\log\frac{m_i^2}{Q^2}-1\right) \right]_{h=(v_1,v_2)} }\\
 & \\
m_2^2= & {\displaystyle
-m_3^2\cot\beta+\frac{m_Z^2}{2}\cos 2\beta
-\sum_i\frac{n_i}{64\pi^2}
\left[ \frac{\partial m_i^2}{\partial h_2}\frac{m_i^2}{h_2}
\left(\log\frac{m_i^2}{Q^2}-1\right) \right]_{h=(v_1,v_2)} }
\end{array}
\ee
while $m_3^2$ can be traded in favour of the one-loop corrected
squared mass $m_A^2$ of the CP-odd neutral Higgs boson, 
as~\cite{ERZ}
\be
\begin{array}{l}
\label{m3}
m_3^2= -m_A^2\sin\beta\cos\beta\\  \\
- \left.{\displaystyle \frac{3g^2m_t^2\mu
A_t}{32\pi^2m_W^2 \sin^2\beta} 
\frac{m_{\;\widetilde{t}_1}^2
\left[\log(m_{\;\widetilde{t}_1}^2/Q^2)-1\right]
-m_{\;\widetilde{t}_2}^2
\left[\log(m_{\;\widetilde{t}_2}^2/Q^2)-1\right]}
{m_{\;\widetilde{t}_1}^2-m_{\;\widetilde{t}_2}^2} }
\right|_{h=(v_1,v_2)}  
\end{array}
\ee

The energy functional corresponding to the ansatz of 
Eq.~(\ref{background}) can be written as:
\be
\begin{array}{rl}
\label{energia}
E_{\rm static}= & {\displaystyle
4\pi\int dr\left\{ \frac{4}{g^2}\left[ (\partial_r\,f)^2
+\frac{2}{r^2}f^2(1-f)^2 \right]  \right. }  \\
&+ {\displaystyle
r^2\left[(\partial_r\,h_1)^2+\frac{2}{r^2}h_1^2(1-f)^2\right]
+r^2\left[(\partial_r\,h_2)^2+\frac{2}{r^2}h_2^2(1-f)^2\right]
}
\\
& {\displaystyle \left.
\phantom{\frac{4}{g^2}}\hspace{-0.6cm}+r^2 
\left[V_{\rm eff}(h_1,h_2)-V_{\rm eff}(v_1,v_2)\right]
\right\} }
\end{array}
\ee
where we have subtracted the vacuum energy. The sphaleron 
is described by the functions $f(r)$, $h_1(r)$ and $h_2(r)$ 
which are the solutions to the equations of motion
\be
\label{h1eq}
\left(\partial_r^2+\frac{2}{r}\partial_r\right)\; h_1(r)
-\frac{2}{r^2}\; h_1(r)\left(1-f(r)\right)^2=\frac{1}{2}
\frac{\partial V_{\rm eff}(h_1,h_2)}{\partial h_1}
\ee
\be
\label{h2eq}
\left(\partial_r^2+\frac{2}{r}\partial_r\right)\; h_2(r)
-\frac{2}{r^2}\; h_2(r)\left(1-f(r)\right)^2=\frac{1}{2}
\frac{\partial V_{\rm eff}(h_1,h_2)}{\partial h_2}
\ee
\be
\label{feq}
\partial_r^2\; f(r)-\frac{2}{r^2}f(r)(1-f(r))(1-2f(r))
=-\frac{g^2}{2}(1-f(r))\left[h_1^2(r)+h_2^2(r)\right]
\ee
subject to the boundary conditions,
\be
\label{boundary}
\begin{array}{l}
h_1(0)=h_2(0)=f(0)=0 \\  \\
h_1(\infty)=v_1,\ h_2(\infty)=v_2,\ f(\infty)=1
\end{array}
\ee
which guarantee that the solutions of (\ref{h1eq}-\ref{feq}) have
finite energy and the correct vacuum behaviour when $r\rightarrow\infty$.

In this section we have introduced the set of supersymmetric
parameters which the sphaleron solutions and sphaleron
energy depend upon. They are $\tan\beta\equiv v_2/v_1$, 
$m_A$, the mass of the CP-odd Higgs
boson, $\mu$ the supersymmetric higgsino mass, $m_Q^2$ and
$m_U^2$, the soft supersymmetry breaking squared mass terms for
the left-handed doublet and right-handed singlet squarks
\be
\label{squarks}
Q_L=\left(
\begin{array}{c}
U_L \\
D_L
\end{array}
\right),\ U_L^c
\ee
and $A_t$ the soft breaking term for the third generation
of squarks corresponding to the tri-linear coupling in the
superpotential (\ref{superp}). 
We will consider $\tan\beta\simlt 15$ and therefore
we can neglect the bottom Yukawa coupling $h_b$, and
thus the corresponding tri-linear soft term $A_b$ will have no
influence on the result, and can be neglected too. As for the
top Yukawa coupling, we will fix $m_t=175$ GeV~\footnote{Notice 
that $m_t$ is the $\overline{\rm MS}$ running top mass
determined at the scale $Q=M_t$. It is related to the pole top
mass $M_t$ by the QCD correcting factor $m_t=M_t/\left[1+4\alpha_s(M_t)
/3\pi\right]$. The on-shell mass $m_t=175$ GeV corresponds to 
$M_t=183$ GeV.} throughout this paper. This value of $m_t$ is
inside the experimental range~\cite{D0,CDF} and we have decided
to fix it in view of the large number of parameters in the
model. Variations of the results in this paper corresponding to
variations of $m_t$ inside the experimental error band 
will not dramatically modify the final conclusions.

\begin{figure}[p]
\centerline{
\psfig{figure=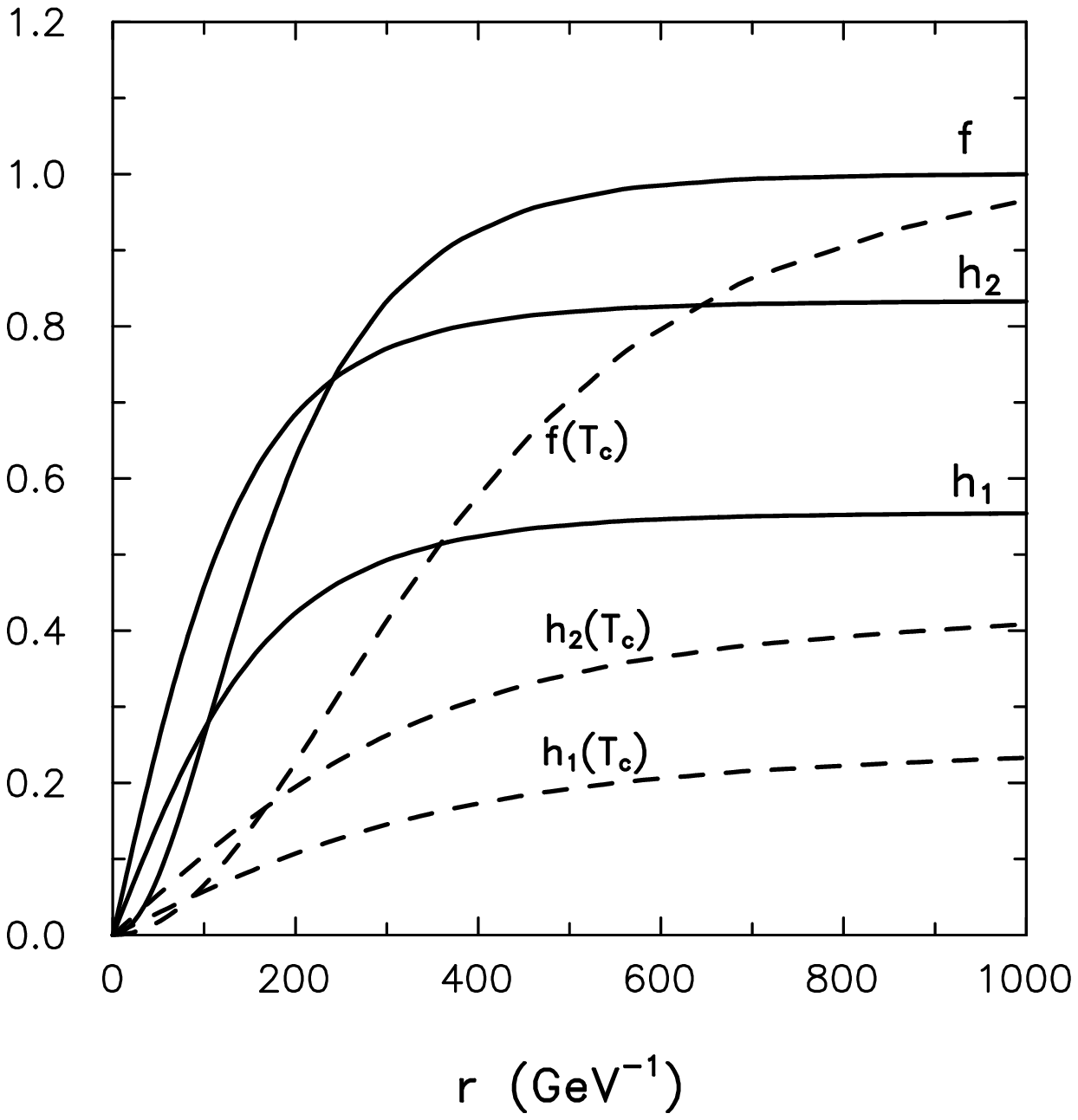,height=13cm,bbllx=6.5cm,bblly=.cm,bburx=16.cm,bbury=13cm}}
\caption{Plots of $h_1(r)/v$, $h_2(r)/v$ and $f(r)$ at zero (solid
lines) and the critical temperature $T=T_c$ (dashed lines) for
$m_t=175$ GeV and
the values of supersymmetric parameters: $\tan\beta=1.5$,
$m_A=100$ GeV, $m_Q=500$ GeV, $m_U=0$ and $A_t=\mu=0$.}
\label{f0}
\end{figure}

\begin{figure}
\centerline{
\psfig{figure=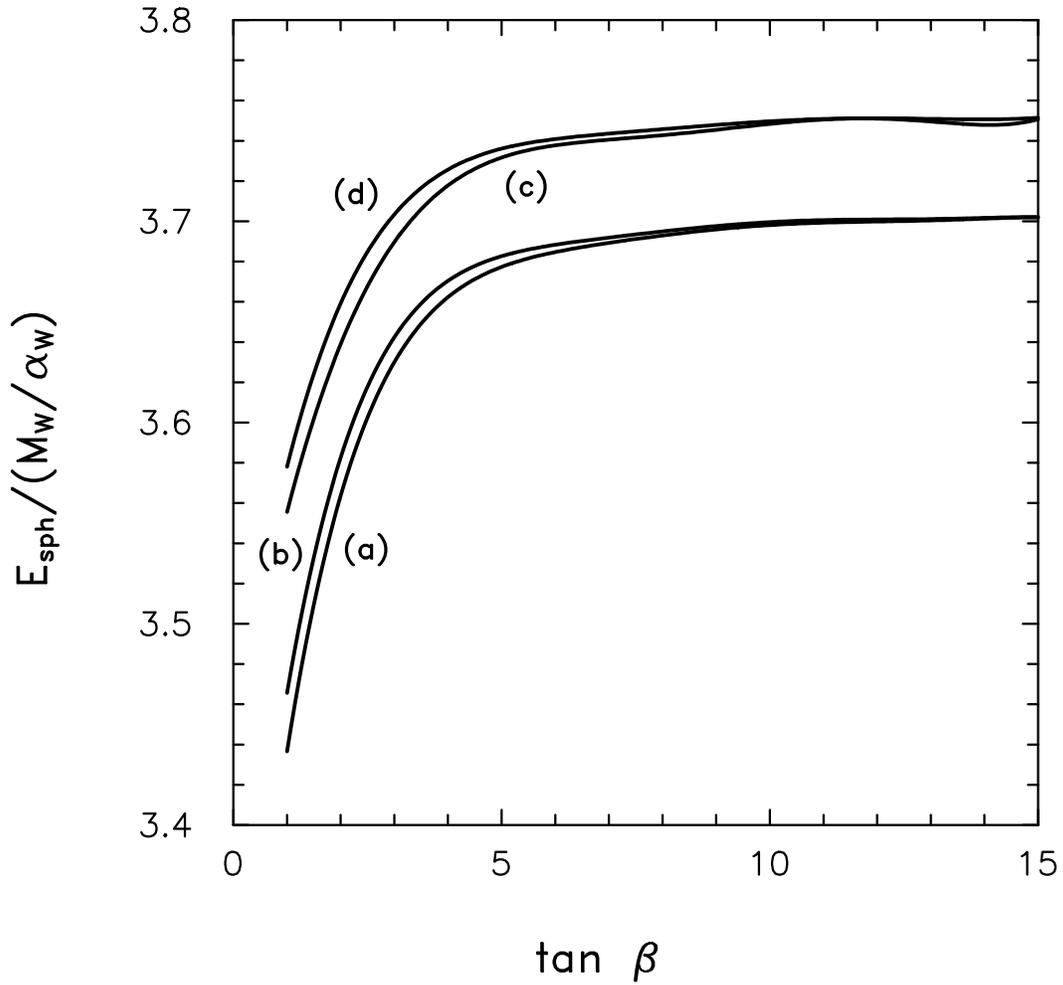,height=13cm,bbllx=6.5cm,bblly=.cm,bburx=16.cm,bbury=13cm}}
\caption{Plot of $\esphal$ as a function of $\tan\beta$ for $m_Q=500$
GeV, $A_t=\mu=0$ and: a) $m_A=100$ GeV, $m_U=0$; b) $m_A=500$ GeV,
$m_U=0$; c) $m_A=100$ GeV, $m_U=400$ GeV; and, d) $m_A=500$ GeV,
$m_U=400$ GeV.}
\label{f1}
\end{figure}

In Fig.~\ref{f0} we have illustrated the functions $h_1(r)/v$,
$h_2(r)/v$ and $f(r)$ [solid lines] for a particular value of
supersymmetric parameters: $\tan\beta=1.5$, $m_Q=500$ GeV,
$m_A=100$ GeV and $m_U=A_t=\mu=0$. In Fig.~\ref{f1} we plot the
sphaleron energy given by Eq.~(\ref{energia}) [in units of 
$m_W/\alpha_W$, $\alpha_W=g^2/4\pi$] as a function of
$\tan\beta$ for different values of the supersymmetric
parameters, where we have varied $m_U=0-400$ GeV and
$m_A=100-500$ GeV. Comparison of curves (a) and (b), and (c)
and (d), shows that the variation of $\esphal$ with $m_A$, within
the considered range, is tiny, while comparison of curves (a)
and (c), and (b) and (d), exhibits the influence of the
parameter $m_U$ in the value of $\esphal$, being $\simlt 3$\%. 
We can see from Fig.~\ref{f1} that
$\esphal$ varies $\sim10$\% over the whole range of
$\tan\beta$. 

\begin{figure}
\centerline{
\psfig{figure=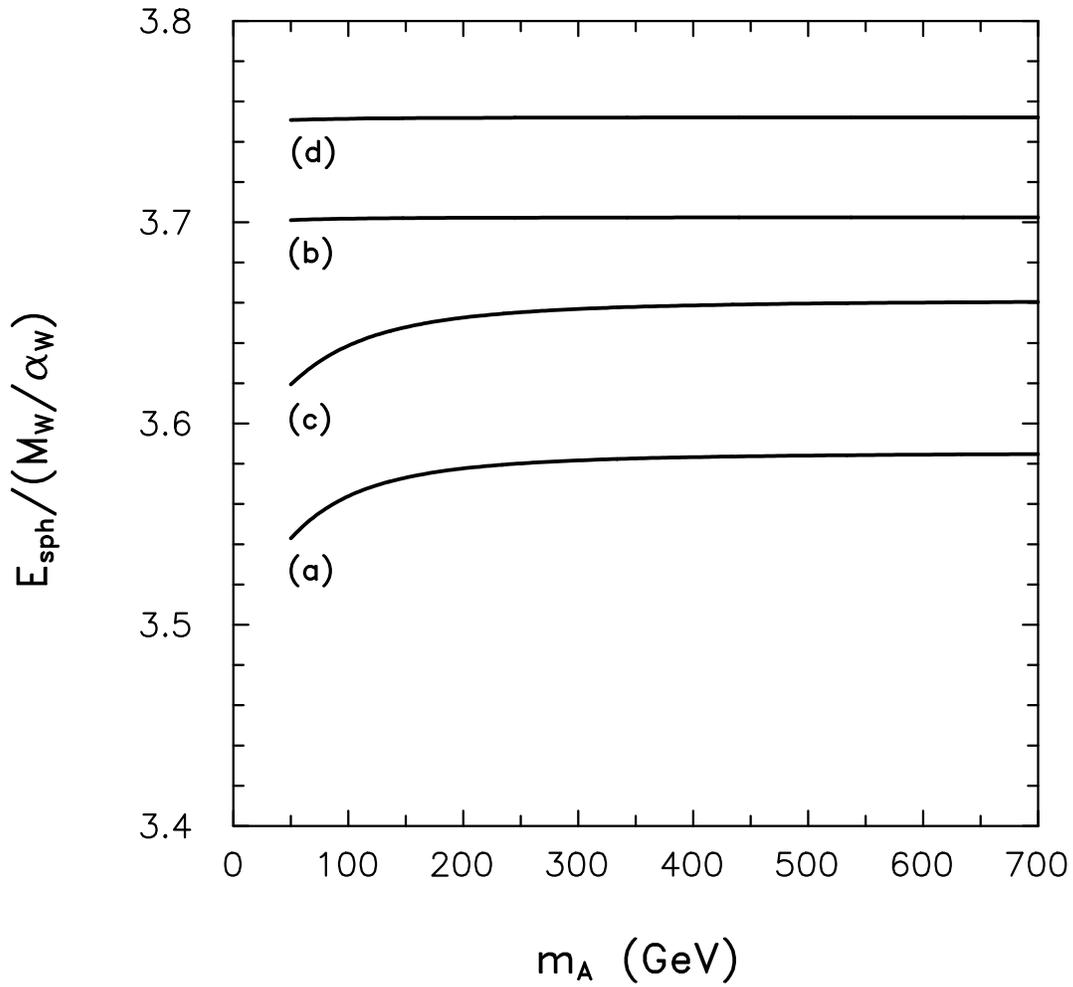,height=13cm,bbllx=6.5cm,bblly=.cm,bburx=16.cm,bbury=13cm}}
\caption{Plot of $\esphal$ as a function of $m_A$ for $m_Q=500$
GeV, $A_t=\mu=0$ and: a) $\tan\beta=2$, $m_U=0$; b) $\tan\beta=15$,
$m_U=0$; c) $\tan\beta=2$, $m_U=400$ GeV; and, d) $\tan\beta=15$,
$m_U=400$ GeV.}
\label{f3}
\end{figure}

A similar exercise to that performed in Fig.~\ref{f1}  has been
done in Fig.~\ref{f3}  exchanging the variable $\tan\beta$ with
the variable $m_A$. In Fig.~\ref{f3} we plot $\esphal$ as a function of
$m_A$ for different values of the other supersymmetric
parameters where we have varied $\tan\beta=2-15$ and $m_U=0-
400$ GeV. The large variation with $\tan\beta$ is explicit by
comparing curves (a) and (b), and (c) and (d). The variation
with $m_U$, comparison of curves (a) and (c), and (b) and (d),
being smaller. Finally the variation with $m_A$ is tiny, as
previously noticed, for the considered range of supersymmetric
parameters. 

\begin{figure}
\centerline{
\psfig{figure=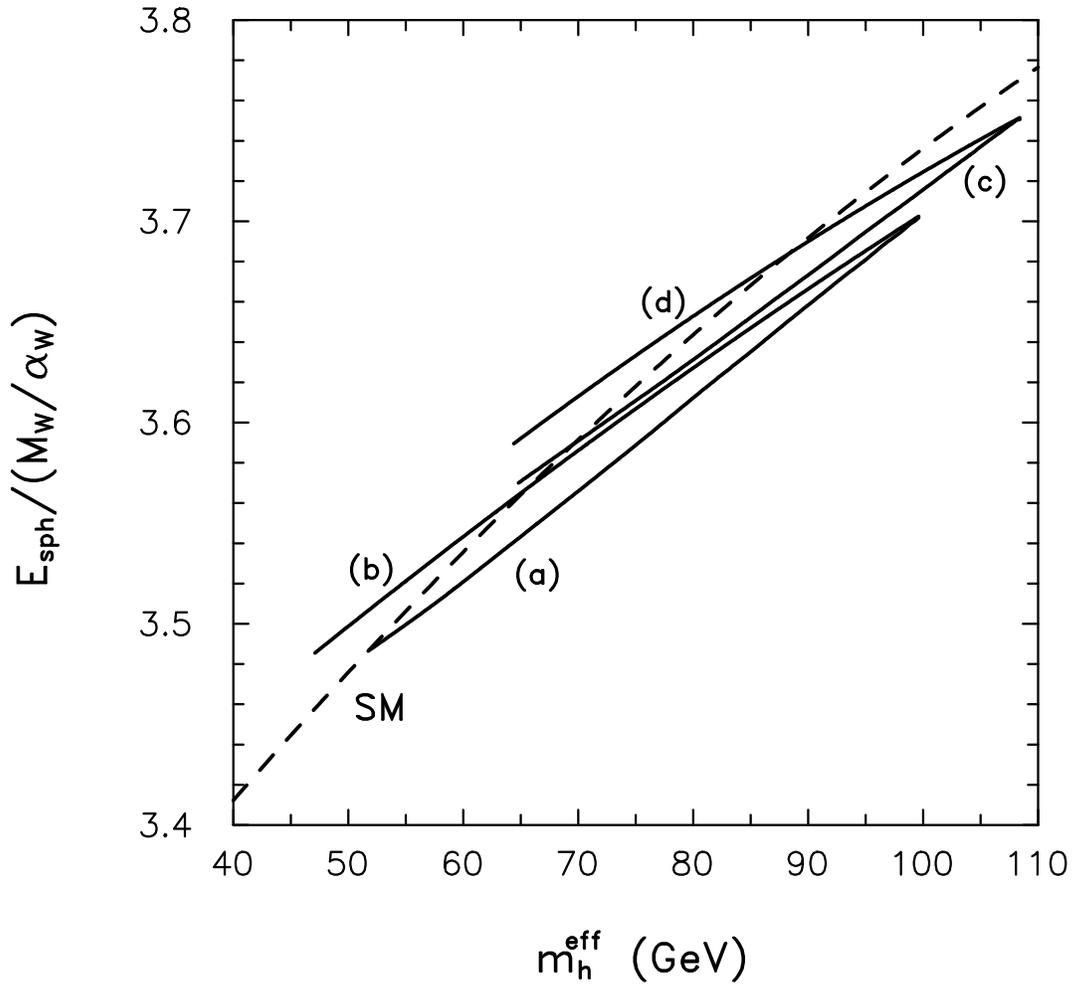,height=13cm,bbllx=6.5cm,bblly=.cm,bburx=16.cm,bbury=13cm}}
\caption{The same as in Fig.~2 but as a function of the effective
Higgs boson mass, $m_h^{\rm eff}$. The dashed line is $\esphal$ for the
Standard Model with a Higgs boson with mass $m_h^{\rm eff}$.}
\label{f2}
\end{figure}

The behaviour of $\esphal$ in Figs.~\ref{f1} and \ref{f3} as a
function of the different supersymmetric parameters can be
understood in a simple way. The Higgs boson mass spectrum
is determined in the MSSM from the tree-level potential
(\ref{tree}) and the radiative corrections (\ref{oneloop}).
In particular the CP-even Higgs boson masses $m_{h,H}$ can be 
determined from the supersymmetric 
parameters~\cite{OYY}-\cite{CEQW}
\be
\label{mh}
m_{h,H}=m_{h,H}(m_A,\tan\beta,m_Q,m_U,A_t,\mu,\dots)
\ee
where the first two parameters determine the tree-level
expression while the rest enter into the radiative corrections.
Now, the field in the $(h_1,h_2)$ plane which acquires
a vacuum expectation value (VEV) is along the direction
$h_1\cos\beta+h_2\sin\beta$, and the 
squared mass (curvature) of
the potential along that direction is
\be
\label{mhefff}
\left(m_h^{\rm eff}\right)^2=\sin^2(\alpha-\beta)\; m_h^2
+\cos^2(\alpha-\beta)\; m_H^2
\ee
where $\alpha$ is the mixing angle in the Higgs sector, 
whose determination, as for the masses $m_h$ and $m_H$
in (\ref{mh}), involves radiative corrections.  The simplest way
of evaluating (\ref{mhefff}), with radiative
corrections provided by (\ref{oneloop}), is by observing that
it does not depend on $m_A$, and therefore, using
the limit (when $m_A\rightarrow\infty$), 
$\sin^2(\alpha-\beta)\rightarrow 1$, one obtains
\be
\label{mheff}
m_h^{\rm eff}=\lim_{m_A\rightarrow\infty}m_h\ .
\ee
Since the sphaleron solution asymptotically points towards
$h_1\cos\beta+h_2\sin\beta$ one expects that
$\esphal$ will mainly depend on 
$m_h^{\rm eff}$~\footnote{We thank J.R.~Espinosa 
for making this observation to us.}. Notice that,
from (\ref{mheff}), $m_h^{\rm eff}$
and $m_h$ will coincide only for very large values 
of $m_A$. This behaviour is exhibited in Fig.~\ref{f2}
where we plot $\esphal$ as a function of $m_h^{\rm eff}$
for the same cases as in Fig.~\ref{f1}, and compare it
with $\esphal^{\rm SM}$ for $m_h^{\rm SM}=m_h^{\rm eff}$. 
We can see that the spreading 
is small ($\sim 1$\%) and due to the fact that the direction
which does not acquire a VEV also contributes to the
sphaleron functional.
 
\begin{figure}
\centerline{
\psfig{figure=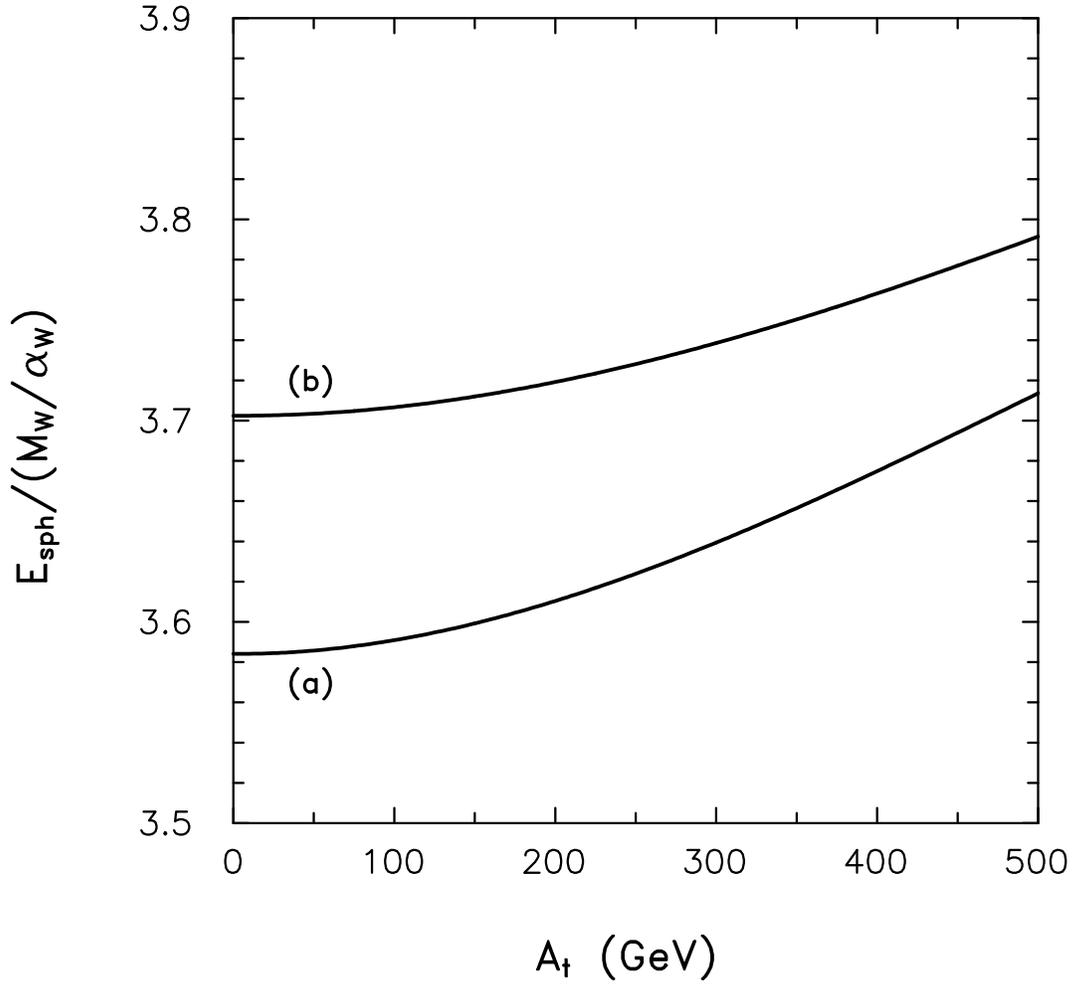,height=13cm,bbllx=6.5cm,bblly=.cm,bburx=16.cm,bbury=13cm}}
\caption{Plot of $\esphal$ as a function of $A_t$ for $m_A=m_Q=500$
GeV, $\mu=m_U=0$, and: a) $\tan\beta=2$; b) $\tan\beta=15$.}
\label{f5}
\end{figure}

\begin{figure}
\centerline{
\psfig{figure=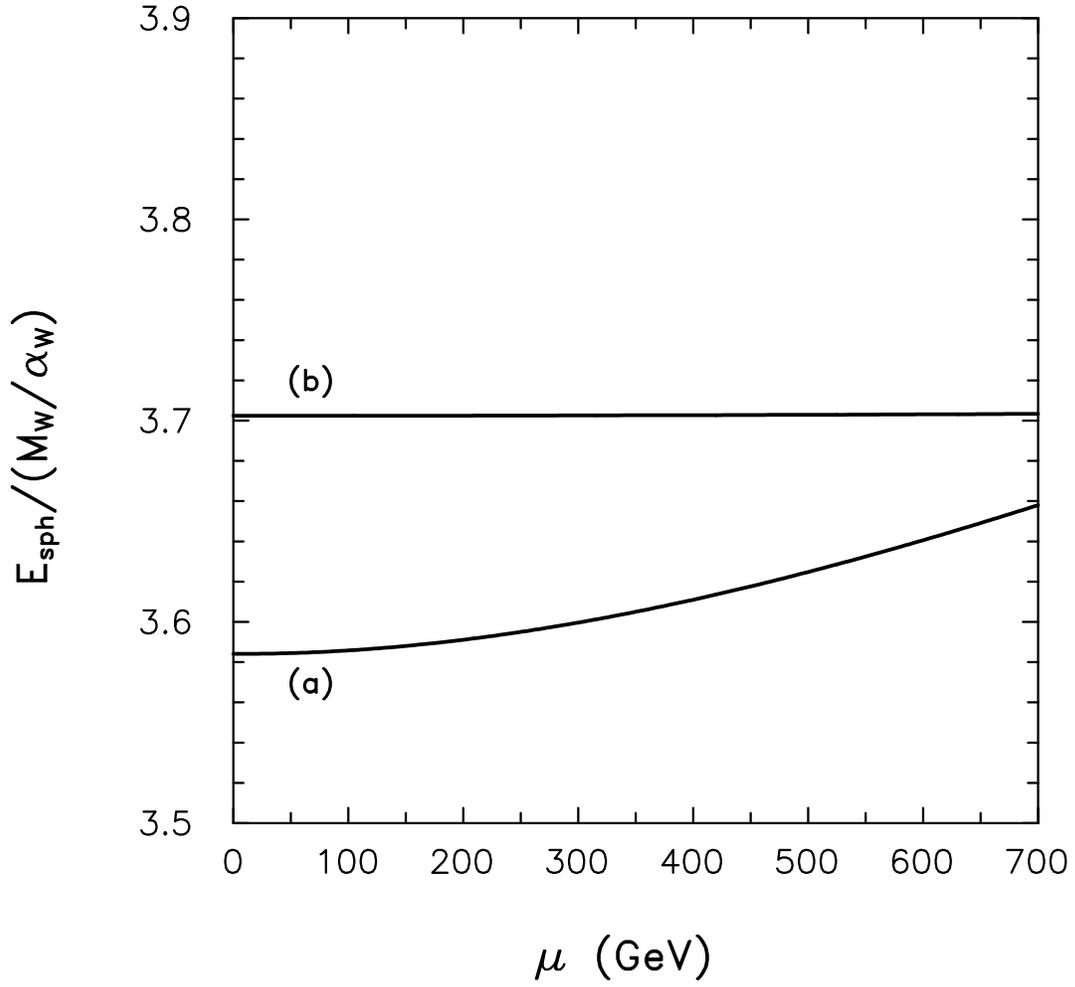,height=13cm,bbllx=6.5cm,bblly=.cm,bburx=16.cm,bbury=13cm}}
\caption{Plot of $\esphal$ as a function of $\mu$ for $m_A=m_Q=500$
GeV, $A_t=m_U=0$, and: a) $\tan\beta=2$; b) $\tan\beta=15$.}
\label{f6}
\end{figure}

Finally the dependence of $\esphal$ with the stop mixing
parameters in Eq.~(\ref{stopentr}), $A_t$ and $\mu$, are shown
in Figs.~\ref{f5} and \ref{f6}. We can see from 
Fig.~\ref{f5} that the variation of
$\esphal$ with $A_t$ is $\simlt 3$\% for generic values of the
supersymmetric parameters, while the variation with $\mu$ is
smaller, as can be seen from Fig.~\ref{f6}. This fact can be 
understood from the fact that the effective stop mixing
parameter is provided by
\be
\label{mixeff}
\widetilde{A}_t=A_t+\mu/\tan\beta
\ee
so that the effect of the $\mu$ parameter is suppressed by
$1/\tan\beta$. For large values of $\tan\beta$, $\esphal$ does
not depend on the $\mu$ parameter, as can be seen from curve (b)
in Fig.~\ref{f6}.

\nsect{Sphalerons at finite temperature}

In order to introduce finite temperature effects we have to
modify the effective potential of Eq.~(\ref{descomp}) by adding
the thermal corrections. We will consistently work at the level
of the one-loop thermal corrections improved by the resummation
of daisy diagrams. The effective potential can be written, to
this level of approximation, as~\cite{mariano2}
\be
\label{descompT}
V_{\rm eff}=V_0(h)+V_1(h)+\Delta V_1(h,T)+\Delta V_{\rm
daisy}(h,T) 
\ee
where $V_0$ and $V_1$ are the tree-level and one-loop potentials
at zero temperature, given by Eqs.~(\ref{treefin}) and
(\ref{oneloop}), respectively, $\Delta V_1$ the one-loop
correction at finite temperature, and $\Delta V_{\rm daisy}$ the
resummation of daisy diagrams. They are given by:
\be
\label{oneloopT}
\Delta V_1(h,T)=\frac{T^4}{2\pi^2}\left\{\sum_i n_i
J_i\left[\frac{m_i^2(h)}{T^2}\right]\right\}
\ee
where $n_i$ is given in Eq.~(\ref{number}), the masses
$m_i(h)$ are defined in Eqs.~(\ref{topmass}-\ref{stopeigen}),
and the thermal function $J_i=J_+(J_-)$, if the $i^{th}$ particle
is a boson (fermion), with
\be
\label{thermalfu}
J_{\pm}(y^2)=\int_0^{\infty}dx\;x^2\log\left(1\mp
e^{-\sqrt{x^2+y^2}} \right).
\ee

As for the term $\Delta V_{\rm daisy}$, it is given by,
\be
\label{daisy}
\Delta V_{\rm daisy}(h,T)=-\frac{T}{12\pi}\sum_B n_B
\left[\overline{m}_B^3(h,T)-m_B^3(h)\right]
\ee
where the sum is extended to scalar bosons and the longitudinal
degrees of freedom of gauge bosons, with $n_B$ given by
Eq.~(\ref{number}) for $\widetilde{\,t}_{1,2}$ and
\be
\label{numberL}
n_{W_{L}}=2,\ n_{Z_L}=1.
\ee
The thermal masses $\overline{m}_B^2(h,T)$ are obtained from
$m_B^2(h)$ by adding the leading $T$ dependent self-energy
contributions, which are proportional to $T^2$. In particular,
the stop squared thermal mass matrix is given by
\be
\label{stopmatT}
\overline{{\cal M}}^2_{\widetilde{t}}=\left(
\begin{array}{cc}
m_{\;\widetilde{t}_L}^2+\Pi_{\widetilde{\;t}_L}(T) 
& m_{\;\widetilde{t}_{LR}}^2 \\
m_{\;\widetilde{t}_{LR}}^2   & m_{\;\widetilde{t}_R}^2
+\Pi_{\widetilde{\;t}_R}(T).
\end{array}
\right)
\ee
The self-energies are given by
\be
\label{pitop}
\begin{array}{rl}
\Pi_{\widetilde{\;t}_L}(T)&=\left[\frac{4}{9}g_s^2+\frac{1}{12}
\left(\Theta_{\st_R}+\sin^2\beta+\cos^2\beta\;\Theta_A
\right)h_t^2
+\frac{1}{4}g^2\right]T^2 \\ & \\
\Pi_{\widetilde{\;t}_R}(T)&=\left[\frac{4}{9}g_s^2+\frac{1}{6}
\left(\Theta_{\st_L}+\sin^2\beta+\cos^2\beta\;\Theta_A
\right)h_t^2
\right]T^2
\end{array}
\ee
where $g_s$ is the strong gauge coupling. Only loops of gauge
bosons, Higgs bosons and third generation quark/squark have
been included, assuming that all the remaining supersymmetric
particles are heavy and decouple by Boltzmann suppression
factors. In particular the gluinos, if light, would provide
contributions to the self-energies which are given by
$2/9\ g_s^2 T^2$. They constitute the main missing
contribution to the thermal masses and would weaken the
strength of the phase transition. We are considering them heavy
as in Ref.~\cite{mariano2}. 
We  also have introduced explicit step-$\Theta$
functions for the contribution of left-handed and right-handed
stops and pseudoscalar bosons, with the convention:
$\Theta_X=1$ [0] for  $m_X\simlt\pi\; T$  
[$m_X\simgt \pi\; T$].

The thermal $W_L$-mass is given by
\be
\label{Wthmass}
\widetilde{m}^{\;2}_{W_L}(h,T)=m_W^2(h)+\Pi_{W_L}(T)
\ee
where the self-energy is
\be
\label{piW}
\Pi_{W_L}(T)=\frac{5}{2}g^2 T^2
\ee
and of course, since we are taking $g'=0$, 
$\widetilde{m}^{\;2}_{Z_L}(h,T)=\widetilde{m}^{\;2}_{W_L}(h,T)$.

Up to now we have considered one-loop corrections to the
effective potential in the sphaleron 
background (ansatz) corresponding to
diagrams with $h_1(r)$ and $h_2(r)$ as external legs. Strictly
speaking we should also consider diagrams with $W_j^a$ as
external legs. They should contribute to the effective action as
\be
\label{accion}
-\frac{1}{2} \Pi_{\mu\nu}^{ab}W_{\mu}^a W_{\nu}^b+\cdots=
-\frac{5}{4}g^2 T^2 W_0^a W_0^a+\cdots\ \ .
\ee
While the leading term in Eq.~(\ref{accion}) cancels for the
sphaleron ansatz of Eq.~(\ref{ansatz}), the ellipsis denotes
terms which are suppressed by powers of the gauge coupling
constant and by inverse powers of the
temperature and we will not consider them explicitly in this 
paper. As previously stated this procedure is self-consistent
since in the MSSM the radiative corrections are dominated by
the top Yukawa coupling and the top/stop sector.


Now, at finite temperature, the energy of the sphaleron, the
equations of motion and the boundary conditions 
for functions $h_1(r,T)$, $h_2(r,T)$ and $f(r,T)$, 
are given by Eqs.~(\ref{energia}-\ref{boundary}), where 
$V_{\rm eff}(h_1,h_2,T)$ is provided by Eq.~(\ref{descompT}) and
$v_1=v_1(T)$, $v_2=v_2(T)$ is the minimum of the
effective potential (\ref{descompT}) at the temperature $T$.
We define the critical temperature $T_c$, as in
Ref.~\cite{mariano2}, as the temperature at which the
determinant of the second derivatives of $V_{\rm eff}(h,T)$ at
$h=0$ vanishes~\footnote{The temperature $T_c$, as defined
in (\ref{critical}), is usually called lower metastability
temperature or lower spinodial decomposition point.  
We thank M. Shaposhnikov for pointing out this to us.
We call it critical temperature just for simplicity.}, i.e.
\be
\label{critical}
\det\left[\frac{\partial^2V_{\rm eff}(h,T_c)}{\partial h_i
\partial h_j}\right]_{h_1=h_2=0}=0 .
\ee
Explicit formulae for computing the critical temperature can be
found in Ref.~\cite{mariano2}. 
For illustrative purposes we have plotted in Fig.~\ref{f0} the
functions $h_1(r,T_c)/v$, $h_2(r,T_c)/v$ and $f(r,T_c)$ 
(dashed lines) for $\tan\beta=1.5$, $m_A=100$ GeV, $m_Q=500$ GeV, 
$m_U=0$ and $A_t=\mu=0$.

\begin{figure}
\centerline{
\psfig{figure=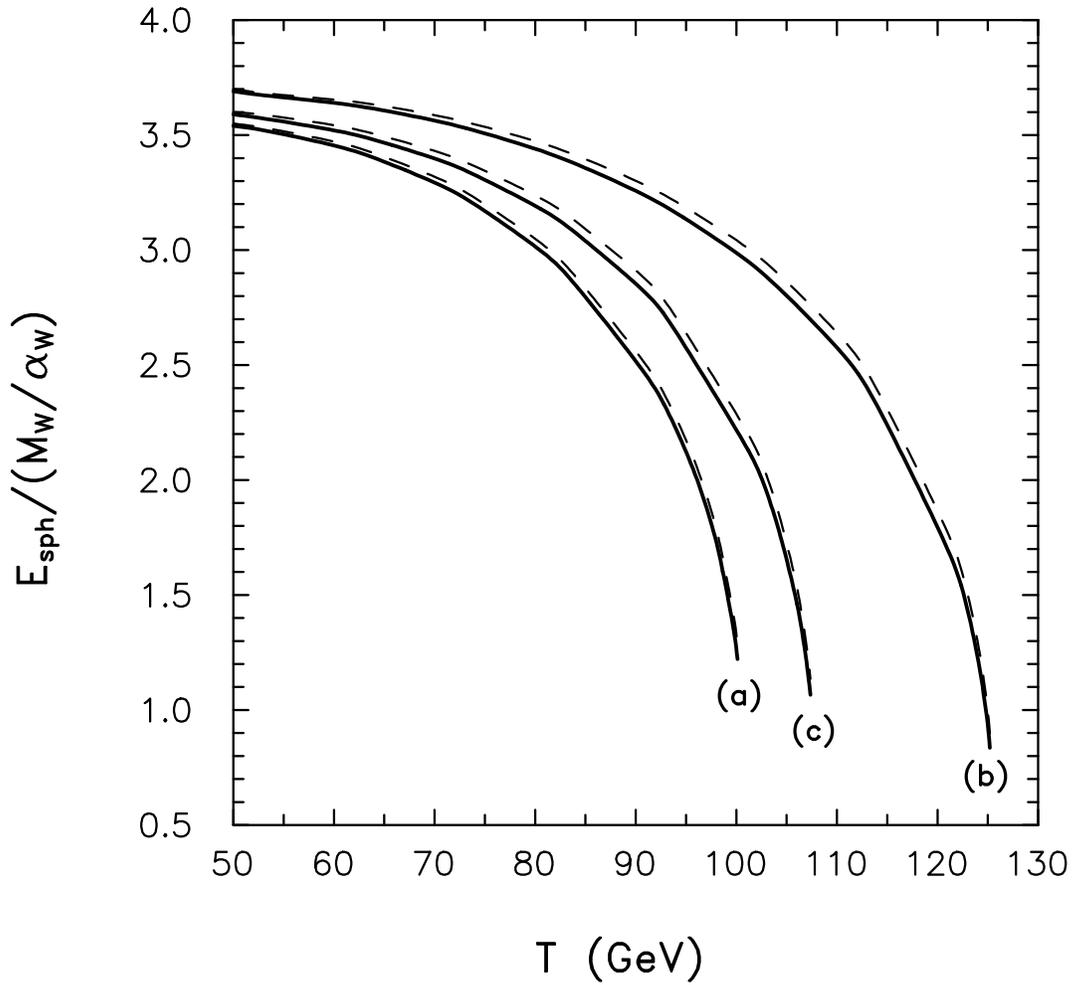,height=13cm,bbllx=6.5cm,bblly=.cm,bburx=16.cm,bbury=13cm}}
\caption{Solid [dashed] lines are plots of $\esphal(T)$ [$\esphal(0)
v(T)/v(0)$] as functions of $T$ for $m_Q=500$ GeV, $\mu=m_U=0$, and:
a) $m_A=100$ GeV, $A_t=0$, $\tan\beta=2$; b) $m_A=100$ GeV, $A_t=0$,
$\tan\beta=15$; and, c) $m_A=500$ GeV, $A_t=200$ GeV, $\tan\beta=2$.}
\label{f7}
\end{figure}

The  $T$-dependence of $\esphal$ is shown in Fig.~\ref{f7} for three
typical sets of supersymmetric parameters (solid lines). In all
cases the pattern of curves is similar to that shown in
Ref.~\cite{Braibant} for the SM case. In particular $\esphal$
decreases with increasing temperatures and
sharply goes to its minimal value at the
temperature where the local minimum we are considering
disappears. We have also probed the approximated scaling law,
\be
\label{scalMSSM}
\esphal^{\rm scal}(T)=\esphal(0)\frac{v(T)}{v}
\ee
where $v(T)=\sqrt{v_1^2(T)+v_2^2(T)}$, $v_1(T)$ and $v_2(T)$
being the vacuum expectation values of the fields $h_1$ and
$h_2$ at finite temperature. We have plotted in Fig.~\ref{f7}
$\esphal^{\rm scal}(T)$ (dashed lines) for the three cases. We
have found that Eq.~(\ref{scalMSSM}) is an excellent
approximation, for all values of supersymmetric parameters used
in Fig.~\ref{f7}, with an error $\simlt 3$\%. In fact we have found that
the scaling law (\ref{scalMSSM}) is not controlled by any
particular combination of supersymmetric parameters, 
but by the strength of the
phase transition. The weaker the phase transition the better
the scaling law. In fact for a second order phase transition  
the scaling law would be exact as happens in the SM. The reason
of the little departure between solid and dashed lines in Fig.~\ref{f7}
is because of the weakness of the phase transition for the
particular values of the chosen parameters. In cases where the
phase transition is stronger, as we will discuss in Section 4,
the departure corresponding to the approximation of
Eq.~(\ref{scalMSSM}) is greater.

\begin{figure}
\centerline{
\psfig{figure=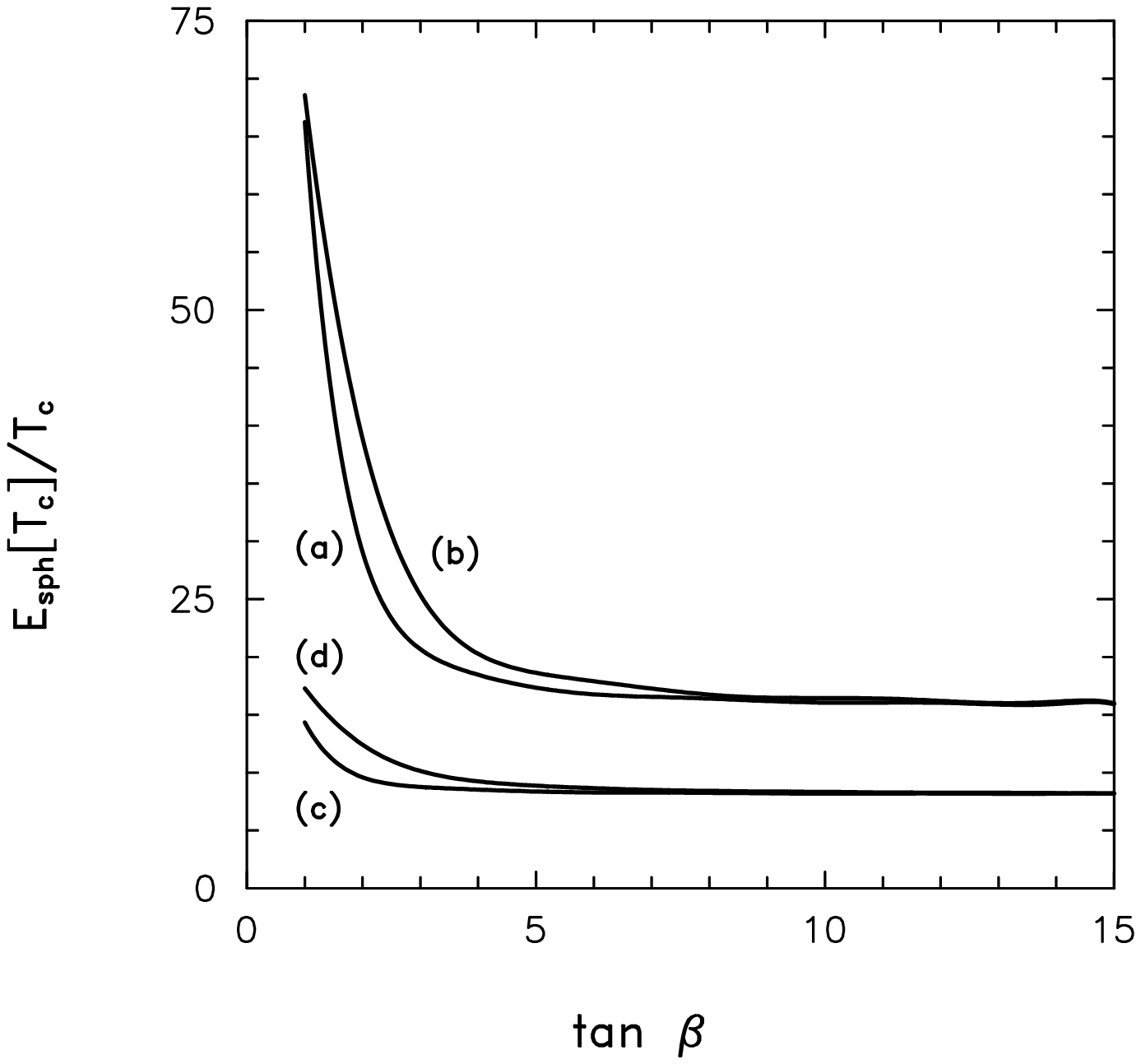,height=13cm,bbllx=6.5cm,bblly=.cm,bburx=16.cm,bbury=13cm}}
\caption{Plot of $\esphal[T_c]/T_c$ as a function of $\tan\beta$ for
the same values of supersymmetric parameters as in Fig.~2.}
\label{f8}
\end{figure}

\begin{figure}
\centerline{
\psfig{figure=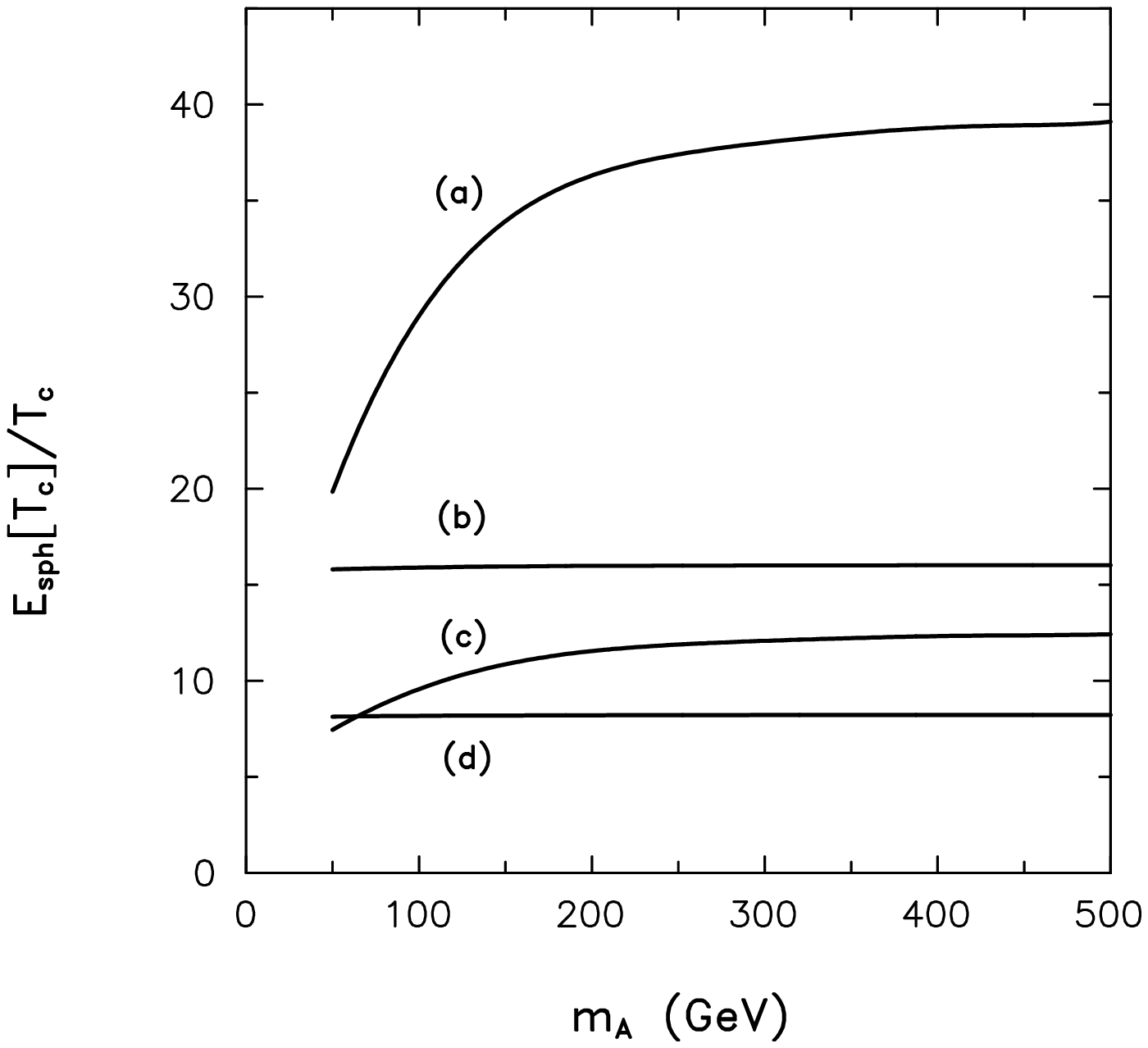,height=13cm,bbllx=6.5cm,bblly=.cm,bburx=16.cm,bbury=13cm}}
\caption{Plot of $\esphal[T_c]/T_c$ as a function of $m_A$ for
the same values of supersymmetric parameters as in Fig.~3.}
\label{f9}
\end{figure}

Finally we have plotted in Figs.~\ref{f8} and \ref{f9}, for the cases 
considered in Figs.~\ref{f1} and \ref{f3}, $\esphal(T_c)/T_c$ as
a function of $\tan\beta$ and $m_A$, respectively. We
can see that the dependence of $\esphal(T_c)$ on the supersymmetric
parameters, is qualitatively similar to that found for
$\esphal(0)$ in Figs.~\ref{f1} and \ref{f3}. 
We can see from Figs.~\ref{f8} and \ref{f9} that for the sample of generic
cases considered up to now the phase transition does not satisfy
condition (\ref{sphalbound})~\footnote{We are making here the 
reasonable assumption that the rate of sphaleron mediated transitions
in the MSSM is provided by~(\ref{sphaltrans}) and, 
therefore, that the effect of the new physics in the MSSM is
entirely encoded in the definition of $\esphal$. Under these
conditions the bound (\ref{sphalbound}) still applies.}
and therefore is not strong enough
first order to generate baryon asymmetry. This problem was taken
care in Ref.~\cite{CQW} where the order of the phase transition
was considered and a `non-generic' region of the
space of supersymmetric parameters was proposed to trigger a
strong first order phase transition. This case is characterized
by light right-handed stops, heavy left-handed stops and low
values of $\tan\beta$ and will be analysed in the next section.

\nsect{The case of light right-handed stops}

In the previous sections we have studied the sphaleron solutions
and energy in the MSSM for generic values of supersymmetric
parameters, as was done in analyses of the phase transition in
the MSSM in the previous
literature~\cite{earlyMSSM}-\cite{mariano2}. Our results on the
value of the order parameter $\esphal(T_c)/T_c$, when compared
with the general bound (\ref{sphalbound}) confirm the former
analyses based on the study of the phase transition: for
generic values of supersymmetric parameters the phase transition
in the MSSM is not strong enough to generate the baryon
asymmetry of the universe. These negative results have motivated
a recent search~\cite{CQW,Delepine} of the region in the
parameter space where the phase
transition has better chances to allow generation of the baryon
asymmetry, while being in agreement with precision electroweak
measurements at LEP~\cite{precision}. 
This region has been identified as follows:
\begin{itemize}
\item
Large values of $m_Q^2$ guarantee small supersymmetric
contributions to the oblique radiative
corrections~\cite{precision}, while small or negative values of 
$m_U^2$ can help in enhancing the value of $R_b$~\cite{Rb}.
\item
Large values of $m_A$ and small values of the stop mixing
parameters $A_t$ and $\mu$ favour the strength of the phase
transition~\cite{mariano1,mariano2}.
\end{itemize}

That region was recently 
analysed in Ref.~\cite{CQW} from the point of
view of the strength of the phase transition, that was found to
be much stronger than in previous analyses. Our aim in this
section is to study the sphaleron solutions and the sphaleron
energy in the MSSM, in the above region of the supersymmetric
parameter space, in order to refine (and eventually confirm) 
the positive results of Ref.~\cite{CQW}. Moreover we will relax
at some point the condition of large $m_A$ 
(even though large values of $m_A$
are favoured by the strength of the phase transition) because
some mechanisms of CP violation associated with supersymmetric
particles require a sizeable variation of $\tan\beta(T_c)$ along
the bubble wall in order to generate the necessary amount of
baryon asymmetry at the electroweak phase transition~\cite{CPMSSM}.

The intuitive explanation of why negative values of $m_U^2$
(keeping the hierarchy $m_Q^2\gg m_U^2$) favour the phase
transition, along with the
appearance of color breaking minima~\cite{CCB} which can endanger the
stability of the standard electroweak minimum,
can be found in Ref.~\cite{CQW}. We refer the
reader, for a thorough explanation of this
phenomenon, to Ref.~\cite{CQW} where all technical details can
be found.

The standard vacuum of the potential (\ref{descomp}), with
radiative corrections provided by (\ref{oneloop}), has a depth
which does not depend on $m_A$, as can be easily checked.
Therefore the simplest procedure is to compute the depth at the
standard vacuum for large values of $m_A$: in this limit the
MSSM goes to the SM with a particular value of the Higgs mass,
determined by the supersymmetric boundary condition of the quartic
coupling,  and the threshold effects due to the stop mixing
parameters, at the scale of supersymmetry breaking ($\sim m_Q$).
These effects can be encoded in an effective quartic
coupling~\cite{CEQW} and a particular value of the SM Higgs
mass, in terms of which the depth of the SM minimum is:
\be
\label{depthSM}
V_{\rm eff}(v)=-\;\frac{1}{4}m_h^2 v^2 .
\ee
On the other hand, for $A_t=0$ and negative values of $m_U^2$
\be
\label{red}
\widetilde{m}_U^2=-m_U^2
\ee
the potential along the direction $U_L^c$ has a minimum
at~\cite{CQW} 
\be
\label{minU}
\langle U_L^c \rangle^2 =\frac{3\;\widetilde{m}_U^2}{g_s^2}
\ee
with a depth
\be
\label{depthU}
V_{\rm eff}(\langle U_L^c
\rangle)=-\frac{3}{2}\frac{\widetilde{m}_U^4}{g_s^2} .
\ee
Comparison of (\ref{depthSM}) with (\ref{depthU}) allows to
determine a critical value of the parameter $\widetilde{m}_U$ 
such that the depth of the charge and color breaking minimum
(\ref{minU}) does not exceed that of the standard electroweak
minimum (\ref{depthSM}):
\be
\label{critU}
m_U^{\rm crit}=\left(\frac{g_s^2}{12} m_h^2 v^2\right)^{1/4}
\ee
For $A_t\neq 0$ we have numerically verified~\cite{CQW}
that the condition $\widetilde{m}_U<m_U^{\rm crit}$ also guarantees
that the standard electroweak minimum is the global minimum
if $A_t\simlt 0.8\; m_Q$. Notice however that such high values of
the mixing parameters are uninteresting for our analysis since
the phase transition will become very weak for those values, as
we will see. 

\begin{figure}
\centerline{
\psfig{figure=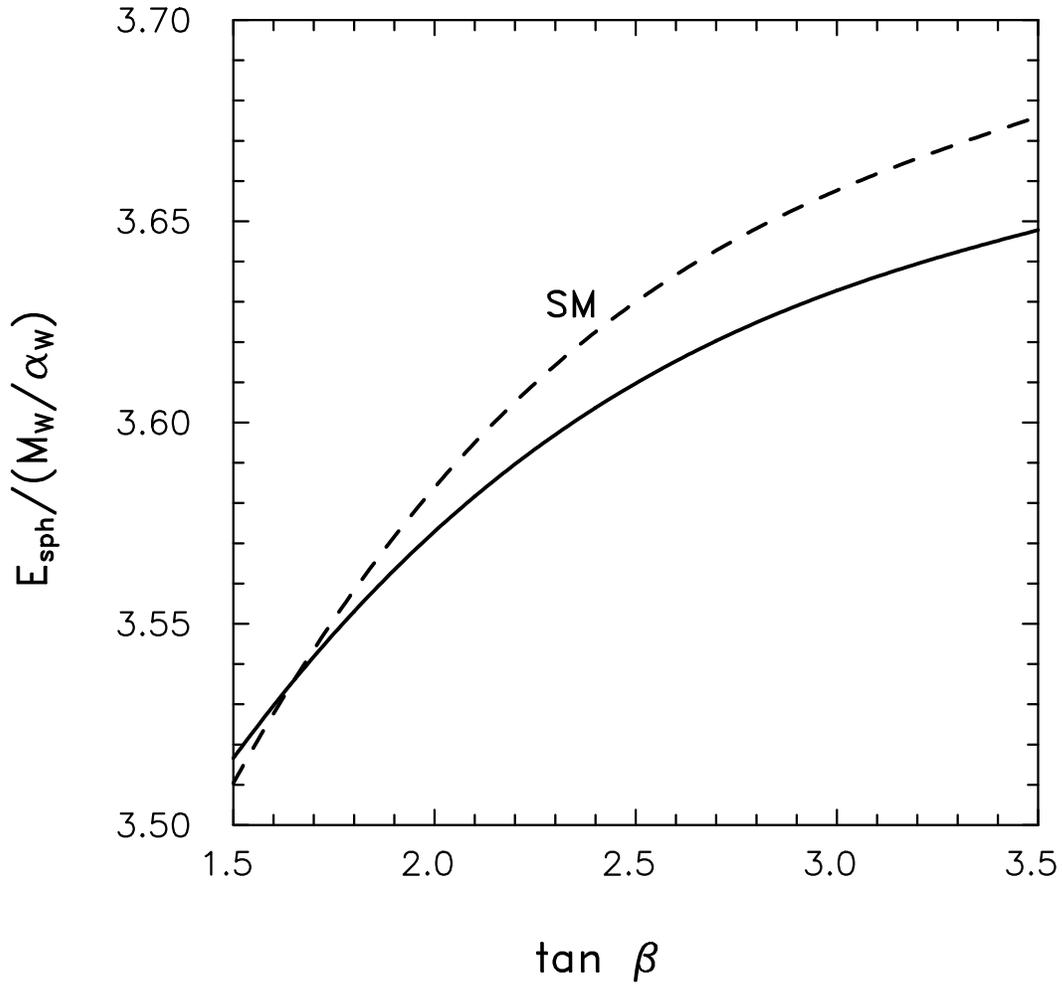,height=13cm,bbllx=6.5cm,bblly=.cm,bburx=16.cm,bbury=13cm}}
\caption{Plot of $\esphal$ at $T=0$ as a function of $\tan\beta$ for
$m_Q=m_A=500$ GeV, $A_t=\mu=0$ and $m_U=m_U^{\rm crit}$. The dashed
line is the Standard Model value for a Higgs mass equal to $m_h^{\rm eff}$.}
\label{f10}
\end{figure}

\begin{figure}
\centerline{
\psfig{figure=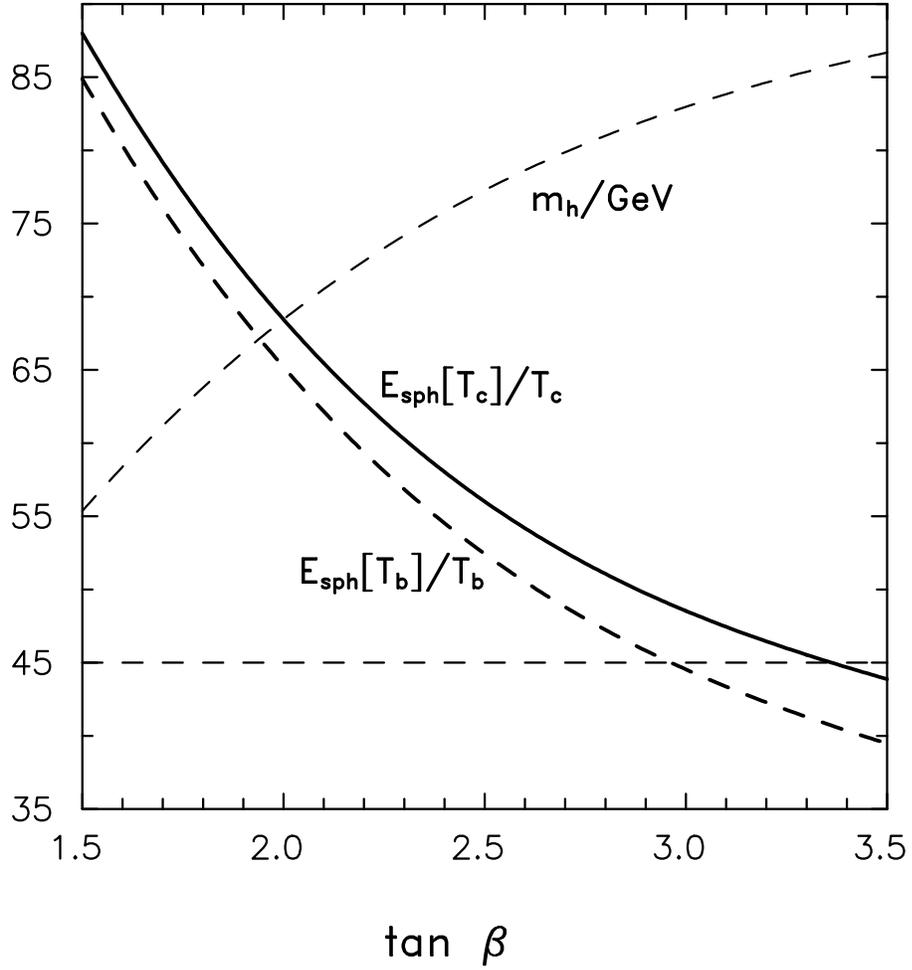,height=13cm,bbllx=6.5cm,bblly=.cm,bburx=16.cm,bbury=13cm}}
\caption{Plot of $\esphal[T]/T$, for $T=T_c,\, T_b$, as function of
$\tan\beta$ for the values of supersymmetric parameters of Fig.~10.
The dashed line is a plot of the lightest Higgs
boson mass for the corresponding values of supersymmetric parameters.}
\label{f12}
\end{figure}

We have then fixed $\widetilde{m}_U=m_U^{\rm crit}$ and 
repeated the analysis of Sections 2 and 3. We will also fix
$m_Q=500$ GeV, to be in good agreement with electroweak
precision measurements, as we said above, and, for the moment,
we will adhere to large values of $m_A$ and low mixing:
$m_A=500$ GeV, $A_t=\mu=0$, although we will relax the latter
conditions later on. For these values of the supersymmetric
parameters we have plotted $\esphal$ at zero temperature as a
function of $\tan\beta$ in Fig.~\ref{f10} [solid line] along with the
comparison, for illustrative purposes, with $\esphal^{\rm SM}$ 
for the value of the Higgs mass given by Eq.~(\ref{mhefff}) 
[dashed line], and
$\esphal(T_c)/T_c$ in Fig.~\ref{f12} [solid line], also as a function of
$\tan\beta$. We also plot in Fig.~\ref{f12} $m_h$ [thin dashed line] as
a function of $\tan\beta$ for the corresponding values of
supersymmetric parameters.

\begin{figure}
\centerline{
\psfig{figure=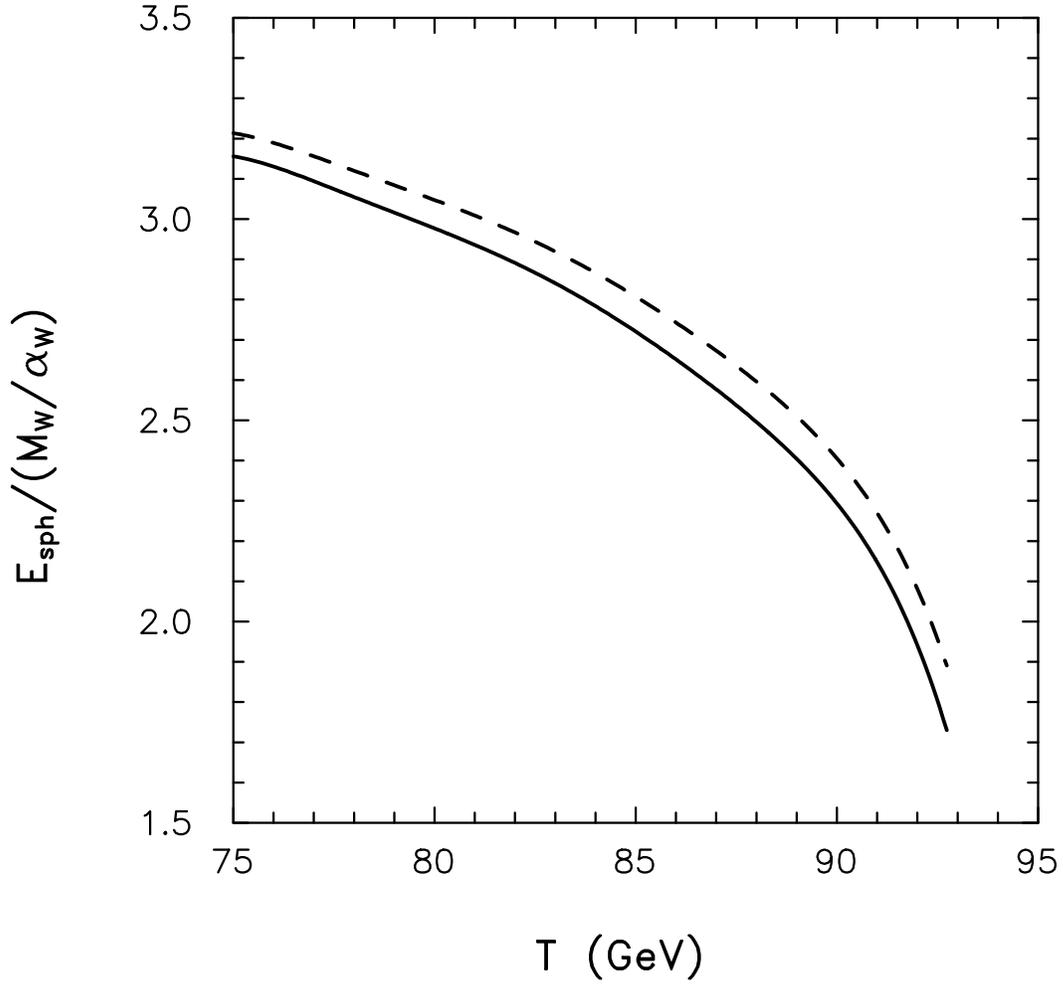,height=13cm,bbllx=6.5cm,bblly=.cm,bburx=16.cm,bbury=13cm}}
\caption{Solid [dashed] line is $\esphal(T)$ [$\esphal(0)
v(T)/v(0)$] as function of $T$ for $\tan\beta=2$ and $m_Q$, $m_A$,
$A_t$, $\mu$ and $m_U$ as in Fig.~10.}
\label{f11}
\end{figure}

Up to here we have used, for our calculations,
the critical temperature $T_c$, where
the origin gets destabilized and the barrier between the origin
and the finite temperature minimum disappears. However tunneling
with formation of bubbles starts somewhat earlier, at a
temperature $T_b$ when the corresponding euclidean action  
$B(T_b)\sim 140$ and the phase transition can proceed
sufficiently fast such that the whole universe can be filled
with bubbles of the new phase. To compute $B(T_b)$ we will use
the analytical estimate obtained in
Refs.~\cite{Carrington,Linde}. First of all we will define an
approximated potential of the form
\be
\label{potapp}
V_{\rm app}=D(T^2-T_c^2)\phi^2-E\;T\phi^3+\frac{\lambda}{4}\phi^4  
\ee
where 
$$
\phi=\sqrt{2}\left(h_1\;\cos\beta+h_2\;\sin\beta\right) ,
$$
and all the constants have been determined numerically. $T_c$
is the critical temperature that we have been using throughout
this paper, $\lambda=m_h^2/(4v^2)$, where $m_h$ is the lightest
Higgs boson mass of the MSSM, 
$$
E=\frac{\lambda}{3}\;\frac{\langle\phi(T_c)\rangle}{T_c}
$$
where $\langle\phi(T_c)\rangle$ is determined numerically, and 
$$
D=\frac{3}{16v^2}\sum_i|n_i|m_i^2.
$$
Secondly, we will use the analytical estimate of
Refs.~\cite{Carrington,Linde} 
\be
\label{fitLinde}
B(T_b)=\frac{38.8
D^{3/2}}{E^2}\left(\frac{\Delta T}{T_b}\right)^{3/2} 
f\left[\frac{2\lambda D}{E^2}\left(\frac{\Delta T}{T_b}
\right)\right]
\ee
where $\Delta T=T_b-T_c$ and the function $f$, defined in
Ref.~\cite{Linde}, equals 1 at zero value of its argument. 
Using (\ref{fitLinde}) one can easily determine the value of the
transition temperature, within the error inherent to the
analytical estimate we are using. This degree of accuracy is
enough for our purposes in this paper. The value of
$\esphal(T_b)/T_b$ is plotted in Fig.~\ref{f12} [thick dashed line].
From Fig.~\ref{f12} we can see that the condition (\ref{sphalbound})
translates into bounds on $\tan\beta$ ($\tan\beta\simlt 3$) and
on $m_h$ ($m_h\simlt 80$ GeV). We can also see that one can
translate the bound (\ref{sphalbound}) at the transition
temperature, into a conservative bound 
at the critical temperature $T_c$, 
\be
\label{fakebound}
\frac{\esphal(T_c)}{T_c}\simgt 50.
\ee
This bound can be useful in many phenomenological studies of
phase transitions. Finally, the goodness of the scaling low
(\ref{scalMSSM}) is shown, for the same values of the supersymmetric
parameters, in Fig.~\ref{f11}, where we see it is accurate with
an error $\simlt 10$\%. 

\begin{figure}
\centerline{
\psfig{figure=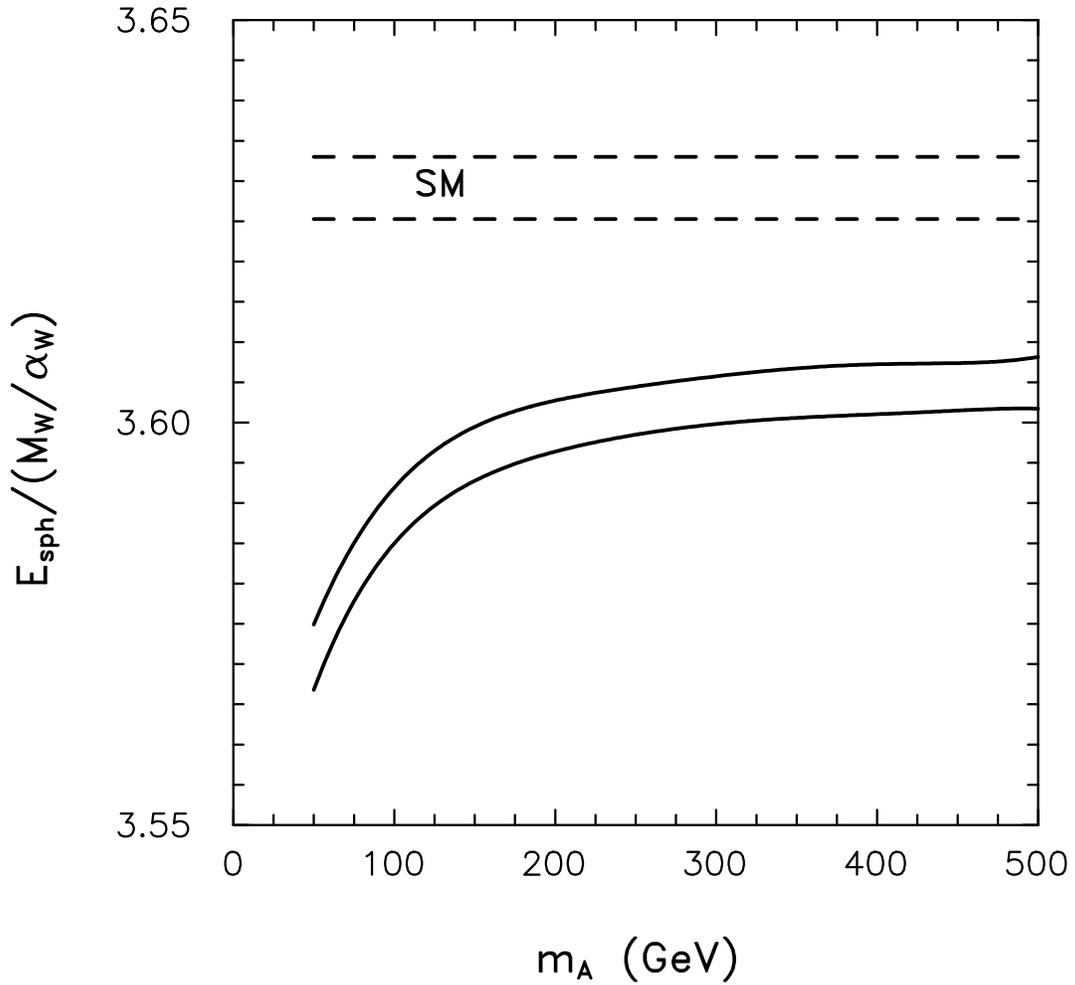,height=13cm,bbllx=6.5cm,bblly=.cm,bburx=16.cm,bbury=13cm}}
\caption{Plots of $\esphal$ at $T=0$ as a function of $m_A$ for
$m_Q=500$ GeV, $A_t=\mu=0$, $\tan\beta=2.4$ (lower solid), 2.5 (upper solid) 
and $m_U=m_U^{\rm crit}$. The dashed
lines are the corresponding Standard Model values for 
a Higgs mass equal to $m_h^{\rm eff}$.}
\label{f14}
\end{figure}

\begin{figure}
\centerline{
\psfig{figure=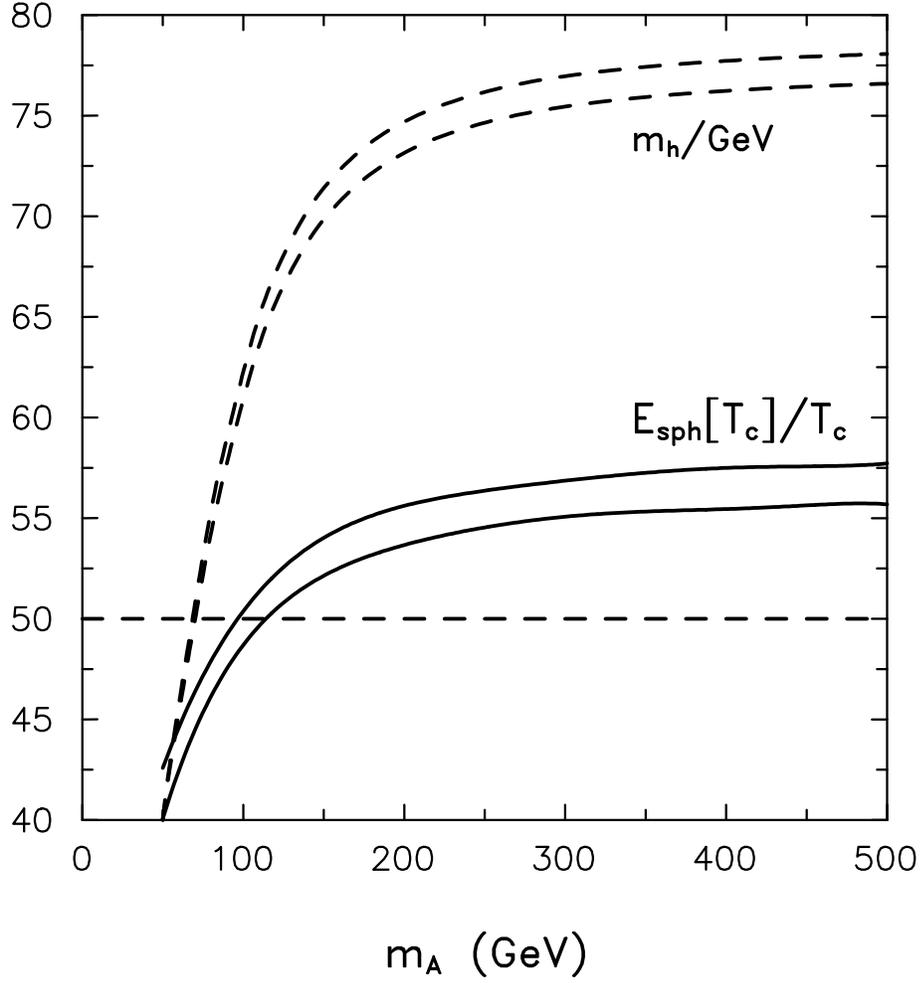,height=13cm,bbllx=6.5cm,bblly=1.cm,bburx=16.cm,bbury=14cm}}
\caption{Plots of $\esphal[T_c]/T_c$, as function of
$m_A$ for the values of supersymmetric parameters of Fig.~13
[upper solid is for $\tan\beta=2.4$ and lower solid for $\tan\beta=2.5$].
The dashed lines are plots of the lightest Higgs
boson mass for the corresponding values of supersymmetric parameters
[upper dashed is for $\tan\beta=2.5$ and lower dashed for $\tan\beta=2.4$].}
\label{f15}
\end{figure}

\begin{figure}
\centerline{
\psfig{figure=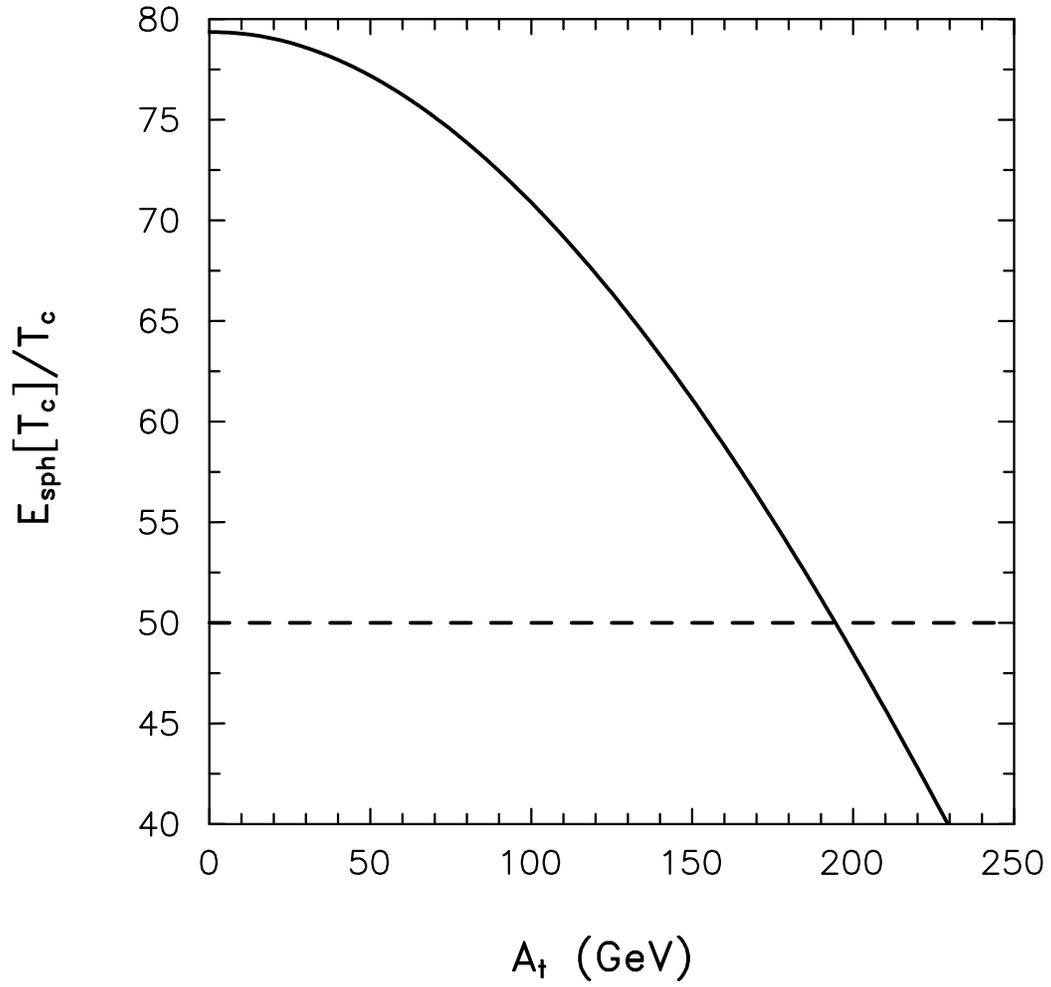,height=13cm,bbllx=6.5cm,bblly=.cm,bburx=16.cm,bbury=13cm}}
\caption{Plot of $\esphal[T_c]/T_c$, as function of
$A_t$ for $\tan\beta=1.7$ and $m_Q$, $m_A$,
$\mu$ and $m_U$ as in Fig.~10.}
\label{f18}
\end{figure}

To obtain the results of Figs.~\ref{f10}-\ref{f11} 
we have been using optimal values of the
supersymmetric parameters, from the point of view of the phase
transition. However, in particular not too large values of
$m_A$ might be required, for baryogenesis purposes,
as we have said above, in order to generate the
needed amount of CP violation at the bubble walls. 
In Figs.~\ref{f14} and \ref{f15} 
we plot $\esphal$ as a function
of $m_A$ for $\tan\beta=2.4,\ 2.5$ and the same values of the other
supersymmetric parameters. In Fig.~\ref{f14} we compare $\esphal$ with
$\esphal^{\rm SM}$ for a Higgs mass $m_h^{\rm eff}$. 
We see they are very close to each other.
In genuine two-Higgs
situations, when $m_A$ is small, they can depart from each other
by an amount $\sim 1.5$\%. In Fig.~\ref{f15} we plot $\esphal(T_c)/T_c$ as a
function of $m_A$. As expected we see that the order parameter
$\esphal(T_c)/T_c$ decreases with increasing $m_A$. Using the bound
(\ref{fakebound}) puts lower bounds on $m_A$: $m_A\simgt 100$ GeV,
for $\tan\beta=2.4$, and $m_A\simgt 110$ GeV for $\tan\beta=2.5$,
the corresponding bounds increasing for larger values of $\tan\beta$.
On the other hand,
LEP results impose bounds in the $(\tan\beta,m_A)$
plane. In particular, for the set of supersymmetric parameters in
Fig.~\ref{f15}~\cite{Janot}: $m_A\simgt 120$ GeV, for $\tan\beta=2.4$,
and $m_A\simgt 110$ GeV, for $\tan\beta=2.5$, the
corresponding bounds decreasing for larger values of
$\tan\beta$~\footnote{We thank P.~Janot for sending us some
original plots with present LEP bounds.}.
We can conclude 
that for $\tan\beta\sim 2.5$, both (\ref{fakebound}) and LEP
results translate into $m_A\simgt 110$ GeV, 
which is a very safe bound for generation of baryon
asymmetry in the MSSM~\cite{CPMSSM}.  
As for the dependence of
$\esphal(T_c)/T_c$ on the stop mixing parameters, we have
plotted in Fig.~\ref{f18} $\esphal(T_c)/T_c$ as a function of $A_t$,
for $\tan\beta=1.7$ and the same values of the other
supersymmetric parameters. We see that, as expected,
$\esphal(T_c)/T_c$ decreases with increasing $A_t$, and the
bound (\ref{fakebound}) puts an upper bound on $A_t$ as 
$A_t\simlt\; 0.4\; m_Q$.

\nsect{Conclusion}

We have constructed the sphaleron solution and computed its energy,
at zero and finite temperature, in the MSSM for an arbitrary set of
supersymmetric parameters. At zero temperature we have included the 
leading one-loop radiative corrections in the presence of the sphaleron. 
At finite temperature we have added the one-loop thermal corrections
where all daisy diagrams are resummed. At zero temperature we have compared
$\esphal$ with the corresponding one for the SM with an effective Higgs mass. 
The difference is small ($\simlt 1.5$\%) even
for small values of $m_A$. We have verified that, at finite temperature, the
scaling law, where all finite temperature effects can be encoded into
the temperature dependence of the vacuum expectation value of the Higgs
fields, is accurate (with an error $\simlt 3$\%), 
for the cases of weak first order phase transition.
For the cases of strong first order phase transition (light stop scenario)
it is somewhat larger ($\simlt 10$\%). Finally our calculation supports the
conclusion that for generic values of 
the supersymmetric parameters the MSSM is
unable to keep any pre-existing baryon asymmetry after the phase 
transition. However we have confirmed the presence of a window for
MSSM baryogenesis corresponding to heavy left-handed stops, light
(lighter than the top) right-handed stops, light CP-even 
higgses ($m_h\simlt 80$ GeV), not so heavy CP-odd higgses 
($m_A\simgt 110$ GeV), moderate stop mixing
($A_t\simlt 0.4\; m_Q$), and small values of $\tan\beta$ 
($\tan\beta\simlt 3$). These results are in qualitative 
agreement~\cite{Jim} with recent non-perturbative calculations
performed in the framework of the MSSM~\cite{CK}, and the
baryogenesis window will be probed at LEP2.

Some unconsidered effects should tend to increase the sphaleron energy,
and thus to improve the previous bounds, while others tend to decrease
it and thus to worsen the bounds. As for the former, 
two loop effects at finite temperature have been recently proved to
strengthen the order of the phase transition in the MSSM~\cite{JR}
and thus they are expected to increase the sphaleron energy.
As for the latter, first of all, we have worked within the
approximation $g'=0$ in order to consider a spherically symmetric ansatz.
${\cal O}(g'^{\; 2})$ effects can be easily accounted for by linearizing the energy
functional with respect to the $U(1)$ gauge field and neglecting the
feed-back on the sphaleron~\cite{Manton}, or by solving explicitly an
axially symmetric ansatz~\cite{angle}: in both cases the SM result for
the value of the sphaleron energy correction is tiny ($\sim 1$\%) as should be 
in the case of the
MSSM. A second effect is characteristic of the nature of the
MSSM due to the large number of scalar fields it contains.
We have set all of them to zero in this work~\footnote{Of course, the
equations of motion for all scalars fields of the theory are 
identically satisfied for zero configurations.}. 
However, in some cases a non-vanishing 
configuration for these fields might minimize the energy functional
leading to a modification of the results contained in this paper. In
any case the variation of $\esphal$ should be negative. 
In particular, for those fields with negligible Yukawa couplings
we expect the energy functional to be minimized by the zero configuration,
since the combined effect of their kinetic terms and the large soft breaking
masses, which are positive definite, will control the configuration of
the energy functional. Only in the case of the third-generation squarks
in Eq.~(\ref{squarks}), and for the case of relatively large values of 
$\widetilde{m}_U$ and/or $A_t$ has the energy functional 
a chance to be minimized for
non-vanishing field configurations, even in the cases of no charge and
color breaking at the vacuum. We are at present investigating this issue
and the results will be published elsewhere~\cite{MOQ}.

\nsect*{Acknowledgements}
We thank M.~Carena, J.~Cline, J.R.~Espinosa, J.-F.~Grivaz, P.~Janot,
M.~Shaposhnikov, C.~Wagner and F.~Zwirner for useful discussions and
comments.

\phantom{.}
\phantom{.}
\phantom{.}
\phantom{.}
\phantom{.}
\phantom{.}
\phantom{.}
\phantom{.}
\phantom{.}
\phantom{.}
\phantom{.}
\phantom{.}
\phantom{.}
\phantom{.}
\phantom{.}
\end{document}